\documentclass[reprint,aps,pra,superscriptaddress,amsmath,amssymb,amsfonts,floatfix,letterpaper]{revtex4-2}

\usepackage{graphicx}
\usepackage{color}
\usepackage{times}
\usepackage[normalem]{ulem}

\usepackage[hypertexnames=false]{hyperref}
\hypersetup{colorlinks=true}
\usepackage[all]{hypcap} 

\newcommand{\beq}{\begin{equation}}
\newcommand{\eeq}{\end{equation}}
\newcommand{\bea}{\begin{eqnarray}}
\newcommand{\eea}{\end{eqnarray}}

\newcommand{\hub}[1]{\textcolor{blue}{\textbf{#1}}}

\begin{document}

\title{Nonequilibrium Quantum Critical Steady State: 
Transport Through a Dissipative Resonant Level}

\affiliation{Department of Physics, Duke University, Durham, North Carolina 27708-0305, U.S.A.} 
\affiliation{Department of Electrophysics, National Chiao-Tung University, HsinChu, Taiwan, R.O.C.}
\affiliation{Physics Division, National Center for Theoretical Sciences, HsinChu, Taiwan, R.O.C.}
\affiliation{Department of Chemistry, North Carolina State University, Raleigh, North Carolina 27695, U.S.A.}
\affiliation{Institute for Quantum Materials and Technologies, Karlsruhe Institute of Technology, 76021 Karlsruhe, Germany}

\author{Gu Zhang}
\affiliation{Department of Physics, Duke University, Durham, North Carolina 27708-0305, U.S.A.}
\affiliation{Institute for Quantum Materials and Technologies, Karlsruhe Institute of Technology, 76021 Karlsruhe, Germany}
\author{Chung-Hou Chung}
\email{chung@mail.nctu.edu.tw}
\affiliation{Department of Electrophysics, National Chiao-Tung University, HsinChu, Taiwan, R.O.C.}
\affiliation{Physics Division, National Center for Theoretical Sciences, HsinChu, Taiwan, R.O.C.}
\author{Chung-Ting Ke}
\affiliation{Department of Physics, Duke University, Durham, North Carolina 27708-0305, U.S.A.}
\author{Chao-Yun Lin}
\affiliation{Department of Electrophysics, National Chiao-Tung University, HsinChu, Taiwan, R.O.C.}
\author{Henok Mebrahtu}
\affiliation{Department of Physics, Duke University, Durham, North Carolina 27708-0305, U.S.A.}
\author{Alex I. Smirnov}
\affiliation{Department of Chemistry, North Carolina State University, Raleigh, North Carolina 27695, U.S.A.}
\author{Gleb Finkelstein}
\email{gleb@phy.duke.edu}
\affiliation{Department of Physics, Duke University, Durham, North Carolina 27708-0305, U.S.A.}
\email{gleb@phy.duke.edu}
\author{Harold U. Baranger}
\email{baranger@phy.duke.edu}
\affiliation{Department of Physics, Duke University, Durham, North Carolina 27708-0305, U.S.A.}
\email{baranger@phy.duke.edu}

\date{26 January 2021}

\begin{abstract}
Nonequilibrium properties of correlated quantum matter are being intensively investigated because of the rich 
interplay between external driving and the many-body correlations. Of particular interest is the nonequilibrium behavior near a quantum critical point (QCP), where the system is delicately balanced between different ground states. We present both an analytical calculation of the nonequilibrium steady-state current in a critical system 
and experimental results to which the theory is compared. The system is a quantum dot coupled to resistive leads: a spinless resonant level interacting with an ohmic dissipative environment. A two channel Kondo-like QCP occurs when the level is on resonance and symmetrically coupled to the leads, conditions achieved by fine tuning using electrostatic gates. We calculate and measure the nonlinear current as a function of bias 
($I$-$V$ curve) 
at the critical values of the gate voltages corresponding to the QCP. 
The quantitative agreement between the experimental data and the theory, with no fitting parameter, is excellent. As our system is fully accessible to both theory and experiment, it provides an ideal setting for addressing nonequilibrium phenomena in correlated quantum matter. 
\end{abstract}

\keywords{\hub{check if we need keywords for PRR}}
\maketitle

\section{Introduction}
\begin{figure}
  \centering
  \vspace*{0.1cm}
     (a)\includegraphics[width=0.31\textwidth]{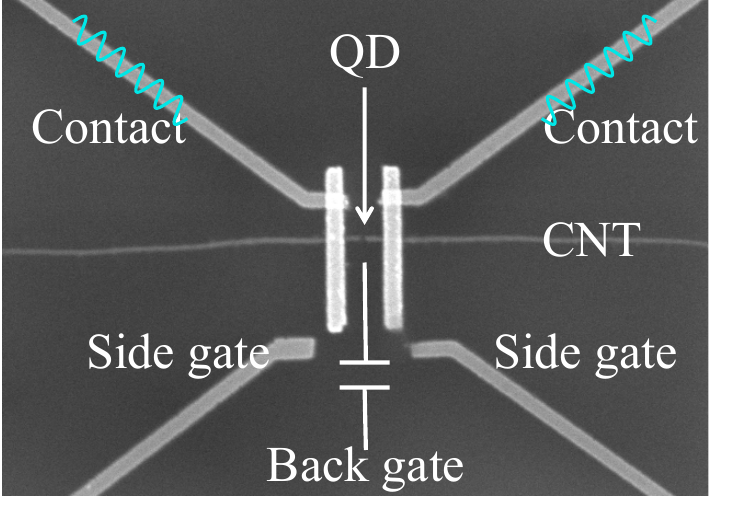}
    \includegraphics[width=0.45\textwidth]{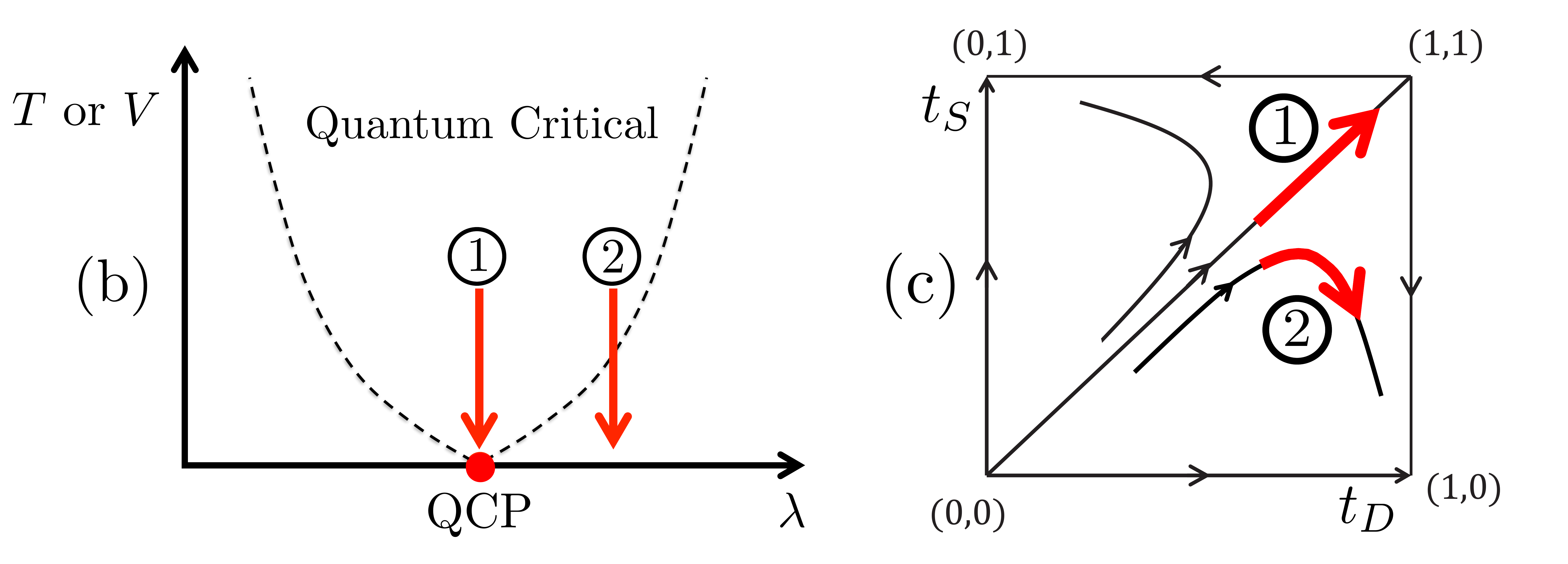}
\vspace*{-0.10in}
\caption{(a) Schematic of the system overlaid on a SEM image of a sample. A quantum dot is formed in the carbon nanotube (CNT) between the source and drain leads. These resistive leads create a dissipative environment for electrons tunneling through the dot. The tunneling barriers can be tuned with the side gates SG1 and SG2. Applying a bias between the source and drain produces a nonequilibrium steady state. 
(b)~Diagram of a quantum critical region as a function of a system parameter, $\lambda$, and temperature or voltage bias. In our case, $\lambda$ could be the dot-leads coupling asymmetry or the dot detuning, tuned by electrostatic gates. When parameters of the system are tuned to their critical values, the quantum critical region extends down to zero temperature or bias (path 1), otherwise a crossover to one of the trivial ground states occurs (path 2). 
(c)~The RG flow of source and drain coupling ($t_S$, $t_D$) when the system is on resonance. For symmetric coupling, the flow is into the full-transmission (strong-coupling) fixed point (1,1) (path 1) which is the QCP. A slight detuning leads to a crossover to a trivial fixed point (path 2). In this work, we focus exclusively on path 1, 
which ends at the QCP.}
  \label{structure}
\end{figure}

Quantum phase transitions (QPT)---abrupt changes of ground state due to quantum fluctuations%
---are of fundamental importance in a wide variety of  many-body systems ranging from quantum materials to quantum magnets and nanostructures \cite{CarrBook,SachdevBook,VojtaPhilMag06}. 
The quantum critical point (QCP) separating the two competing ground states dominates physical properties even at finite temperature where a quantum critical region exists (see Fig.\,\ref{structure}).
By tuning parameters  of the system to their critical values, the system stays in the critical region down to zero temperature, 
such as for path 1 in Fig.\,\ref{structure}. 
In contrast, detuning results in a crossover from quantum critical behavior to one of the trivial phases (path 2).
Along path 1, it is well established that at low temperature thermodynamic observables show universal scaling. 
Properties away from equilibrium, such as when a bias is applied (a nonequilibrium steady state) or a parameter suddenly changed (a quantum quench), are much less well understood. Indeed, quantum nonequilibrium phenomena are receiving increasing attention 
in recent years
\cite{CarrBook,PolkovnikovNoneqClosedRMP11,EisertNoneqreviewNatP15, RylandsNoneqIntegrable20, LeHurDrivenQImpReviewCRP18, NoneqDMFT-RMP14}. 
Here we present a comprehensive theoretical and experimental study of the nonequilibrium steady-state current in a system tuned to criticality.

QPT occur not only in the bulk but also on the boundary of interacting systems, as in quantum impurity models \cite{VojtaPhilMag06, NoneqDMFT-RMP14, SiReviewPSSb13, ChowdhuryIngersentPRB15, Vojta_critquasiEPJST15, LeHurDrivenQImpReviewCRP18, MocaZarandPRL19}.  
The two-channel Kondo model is a prototypical example: two independent metallic channels each screen a localized spin, resulting in frustration and a non-Fermi liquid QCP. 
Nanoscale systems are ideal for studying such 
impurity QPT because of the exquisite control over parameters that they provide 
\cite{DavidGG_CarrBook,FlorensReviewJPCM11,BauerleCoherentRevRPP18, LeHurDrivenQImpReviewCRP18, HartmanFolkEntropyNatPhys18, MocaZarandPRL19, ChoiSim4DotsPRB20}. 
Indeed, a growing number of QPT are being studied in nanosystems, including e.g.\ both spin and charge two channel Kondo systems 
\cite{Potok2CK07,Mebrahtu12,Mebrahtu13,KellerDGGNat15,IftikharPIerreNat15,IftikharPierreScience18,AnthorePierrePRX18}. 
These nanosystem QPT provide insight into more complex quantum impurity QPT, such as 
arise in 
strongly correlated materials \cite{NoneqDMFT-RMP14, SiReviewPSSb13, ChowdhuryIngersentPRB15, HartleMillis15}. 

Here we present both an analytical calculation and a direct measurement of the steady-state 
current as a function of bias voltage ($I$-$V$ curve) 
with system parameters tuned exactly to a QCP. 
The system is a spin-polarized carbon nanotube quantum dot connected to resistive leads via tunable tunnel barriers (Fig.\,\ref{structure}). The resistance of the leads creates an ohmic dissipative environment
\cite{Bomze09,Mebrahtu12,Mebrahtu13}, and the quantum dot serves as the quantum impurity. 
The QCP occurs when both (i) a level in the dot is resonant with the leads and (ii) the dot is symmetrically coupled to them. Both of these properties can be fine-tuned using gate voltages, enabling experimental access to the QCP. At the QCP, which is of the \emph{two channel Kondo type} \cite{Mebrahtu12}, the conductance through the dot 
becomes perfect ($e^2/h$ when $T\!\to\!0$), while otherwise it tends to zero. We previously presented several scaling relations, including non-Fermi liquid scaling along path 1, 
measured as a function of $T$ for negligible bias, i.e.\ in the equilibrium regime \cite{Mebrahtu13}.

We exploit a close connection between the system studied here and resonant tunneling in an interacting one-dimensional (1D) system, a Luttinger liquid (LL). 1D systems have, of course, played a central role in contemporary condensed matter physics \cite{GiamarchiBook,GogolinBook}. Such a connection, which seems surprising at first since there is nothing one-dimensional in our system, derives from the essentially 1D nature of all quantum impurity problems. Indeed, tunneling in a resistive environment can be viewed as a \emph{quantum simulation} of tunneling in a LL with repulsive interactions \cite{MatveevGlazman93,FlensbergPRB93,SassettiWeissEPL94,SafiSaleurPRL04,LeHurLiPRB05, BordaPRB05,*BordaZarandX06, Mebrahtu12,Mebrahtu13,JezouinPierre13,AnthorePierrePRX18, LeHurDrivenQImpReviewCRP18}.
In equilibrium, resonant peaks of perfect conductance in a LL have been extensively studied theoretically   \cite{KaneFisherPRB92,*KaneFisherPRB92a,EggertAffleck92,Furusaki93,*Furusaki98,YiKanePRB98,*YiPRB02,NazarovGlazman03,polyakov03,KomnikGogolinPRL03,Meden2005,GoldsteinPRL10a,HuKaneX16}. 
This system has a similar two channel Kondo-like QCP separating weak-tunneling regimes dominated by one barrier, either source or drain 
\cite{EggertAffleck92,YiKanePRB98,KomnikGogolinPRL03}. 
However, there are significant differences (discussed below) near the QCP. 

The interplay between nonequilibrium and many-body effects has been studied in a variety of nanosystems through nonlinear $I$-$V$ characteristics \cite{AlhassidRMP00, ChangLLrevRMP03, Micolich0.7Rev11, LevyYeyatiSqdotS-rev11, DavidGG_CarrBook, AnthoreUniversalityRevEPJST20}. 
Experimental systems studied include the Kondo effect in quantum dots, tunneling into edge channels, and dissipative tunneling. 
However, to our knowledge, the nonequilibrium $I$-$V$ curve of a system tuned exactly to 
the critical value of the QCP control parameter has not been measured previously, except for  preliminary indications in our own work \cite{Mebrahtu13, PathTwo}.
Nonlinear $I$-$V$ curves in the critical regime of a QCP have certainly been reported (for example \cite{LeturcqEnsslinPRL05,Makarovski07b,Potok2CK07,ChangMarcusPRL13,KellerDGGNat15}), but these all involved measurements through tunnel barriers that therefore probe the density of states in 
the system \emph{in equilibrium}. Such a measurement is tunneling or transport spectroscopy \cite{[{}][{, pp.\ 199, 310-4.}]{IhnBook}}, 
in which no truly nonequilibrium effects are involved \cite{PustilnikPRB04}. 
Theoretically, 
in the scaling regime in which $I\propto V^\alpha$, the exponent $\alpha$ has been frequently deduced from the scaling dimension of the leading operators at the QCP (see for example \cite{FisherGlazman97,AffleckLesHouches10}). 
Nonequilibrium properties at the LL resonant tunneling QCP, for instance, have been studied only in this way, thus capturing only the exponent in the scaling regime (see e.g.\,\cite{KaneFisherPRB92,EggertAffleck92}).  
A few full calculations beyond the scaling exponent exist in the literature. First, approximate numerical treatments have been employed \cite{HettlerPRL94,*HettlerPRB98,vonDelftAnPhys99,BuxboimPRB03,KirchnerSiPRL09, LeeChungPRB13, LandauCornfeldSelaPRL18}, though not of the model we study. Second,  
analytical $I$-$V$ curves have been obtained for the crossover from a QCP to a Fermi liquid state for the two-impurity, two-channel, and topological Kondo models \cite{SelaExactTransPRL09,*SelaNoneqQdotsPRB09,MitchellSelaPRL16,BeriPRL17}.
In Fig.\,\ref{structure}(c) these correspond to properties along 
the vertical line from point (1,1) to (1,0) 
in which one moves out of the quantum critical region. In contrast, here we are 
interested in paths that 
end at the QCP, such as path 1.
We see, then, that the nonequilibrium $I$-$V$ characteristics of a system tuned to a QCP have received surprisingly little attention, either experimentally or theoretically. 

In this work, we rectify this oversight. 
Experimentally, a key feature of our samples is sufficient gates to allow tuning of the system to have both resonant transmission and equal tunneling barriers between the carbon nanotube quantum dot and the leads (see Fig.\,\ref{structure}). 
Theoretically, we approach the problem using a field-theoretic description and bosonization \cite{GiamarchiBook,GogolinBook}. 
The key steps in our argument are to use universality arguments to find the form of the effective Hamiltonian at the QCP and then to incorporate the voltage bias into it. A Keldysh calculation to second order in the deviation from the QCP but to all orders in the coupling to the environment then yields the $I$-$V$ curve in the universal regime.  

The paper is organized as follows. 
We start by defining our model for the dissipative resonant-level problem (Sec.\,\ref{sec:model}) 
and then indicate the key steps to obtain 
the effective Hamiltonian at the QCP, Eq.\,\eqref{eq:Hboson} (Secs.\,\ref{sec:qflucts}-\ref{sec:qc_regime}). 
We find the $I$-$V$ characteristics in the quantum critical regime beyond simple scaling in Sec.\,\ref{sec:I-Vcurve}. 
The main theoretical result of the paper is the analytical expression for the nonlinear $I$-$V$ curve at finite temperature, Eq.\,\eqref{eq:full_I-V}. 
Experimental results are then presented and compared with the theory (Sec.\,\ref{sec:experiment}): Fig.\,\ref{fig:GV} shows excellent agreement between theory and experiment. 

Sec.\,\ref{sec:DCB} presents an alternative physical understanding of the non-Fermi-liquid quantum critical state in terms of tunneling in an environment (dynamical Coulomb blockade), and our conclusions are in Sec.\,\ref{sec:conclusions}. Several appendices (App.\,\ref{sec:weak}-\ref{sec:derriveDCB})
present details of our argument.  


\section{Model and Hamiltonian}
\label{sec:model}

The system is shown in Fig.\,\ref{structure}: a spinless resonant level between two resistive leads. The Hamiltonian 
consists of several parts, 
\begin{equation}
H=H_\textrm{Dot}+H_\textrm{Leads}+H_\mu+H_\textrm{T}+H_\textrm{Env}.
\label{eq:H}
\end{equation}
$H_\textrm{Dot}\!=\!\epsilon_\textrm{d} d^\dagger d$ 
models the dot with single energy level $\epsilon_\textrm{d}$ which may be tuned by a gate voltage.
We focus on the resonant condition $\epsilon_\textrm{d} \!=\! 0$. 
$H_\textrm{Leads} \!=\! \sum_{\alpha=\textrm{S,D}} \sum_k \epsilon_k c_{k\alpha}^\dagger c_{k\alpha}$
represents non-interacting electrons in the source (S) and drain (D) leads. 

Tunneling in our system excites the resistive environment through fluctuations of the voltage on the source and drain. These require a quantum description of the tunnel junction \cite{IngoldNazarov92,NazarovBlanterBook,DevoretEsteveUrbina95,VoolDevoretIJCTA17} via junction charge and phase fluctuation operators that are conjugate to each other, $\varphi_{S/D}$ and  $Q_{S/D}$. A tunneling event shifts the charge on the corresponding junction, as, for example, in this contribution to tunneling from the dot to the source:  
$c_{kS}^{\dagger}e^{-i\sqrt{2\pi}\varphi_{S}}d$. 
We take the capacitance of the two tunnel junctions to be the same and so it is natural to consider the sum and difference variables 
$\psi \equiv (\varphi_S + \varphi_D)/2$ and 
$\varphi \equiv \varphi_S - \varphi_D$.
The fluctuations $\varphi$ involve charge flow through the system and so couple to the environment. In contrast, because $\psi$ involves the total dot charge, it is not coupled to the environment \cite{IngoldNazarov92,NazarovBlanterBook,LiuRLdissipPRB14}, and we therefore drop it at this point. 
We thus arrive at the tunnel Hamiltonian
\begin{equation}
H_\textrm{T}= \sum_k \left( t_S c_{kS}^{\dagger}e^{-i\sqrt{\frac{\pi}{2}} \varphi}d
+t_D c_{kD}^{\dagger}e^{i\sqrt{\frac{\pi}{2}} \varphi}d+{\rm h.c.} \right).
\label{eq:HT}
\end{equation} 
The barriers to tunneling are large in the experimental system and so  $t_S$ and $t_D$ are small---the system is in the ``weak coupling'' regime. We focus on the symmetric case $t_S = t_D$.

The ohmic environment of resistance $R$ is modeled in the usual way as a bath of harmonic oscillators to which the phase fluctuations $\varphi$ of the junction are coupled. The model must produce the expected temporal correlations of the phase fluctuations, namely 
$\langle e^{-i\varphi(t)}e^{i\varphi(0)}\rangle \propto (1/t)^{2r}$
where the exponent $r$ is related to the resistance of the environment by 
$r\equiv Re^2/h$ \cite{IngoldNazarov92,NazarovBlanterBook,DevoretEsteveUrbina95,VoolDevoretIJCTA17}. 
We choose to represent the environment by bosonic fields $\varphi(x)$ and its conjugate $\vartheta(x)$ with a transmission line Hamiltonian  
\begin{equation}
H_\textrm{Env} =  \frac{1}{2} \int_0^\infty dx \, \left[ \frac{1}{2r} (\partial_x \varphi)^2 
+ 2r (\partial_x \vartheta)^2 \right] .
\label{eq:Henv}
\end{equation}
It is coupled to the junction by identifying  $\varphi(x\!=\!0)$ as the phase  $\varphi$ in the tunneling term Eq.\,\eqref{eq:HT}. 


Finally, the term driving the system out of equilibrium is  
\begin{equation}
H_\mu= \sum_{\alpha=\textrm{S,D}} \sum_k \mu_\alpha c_{k\alpha}^{\dagger}c_{k\alpha}, 
\label{eq:Hmu}
\end{equation}
where the chemical potential is related to the applied bias, $\mu_{S/D}=\pm eV/2$. 
It is often convenient to handle the bias through a time-dependent gauge transformation \cite{IngoldNazarov92, KaneFisherPRB92} that moves it to the tunneling term: physically, when an electron hops from a lead to the dot it acquires a phase factor corresponding to the change in energy (drop in bias) across that barrier. With symmetric tunneling and capacitance, the bias voltage drops symmetrically, and so each tunneling term acquires a phase factor $e^{\pm ieVt/2}$.

\section{Quantum fluctuations enhance transmission}
\label{sec:qflucts}

A great deal is known about the equilibrium properties of this system \cite{Mebrahtu12,Mebrahtu13,LiuRLdissipPRB14,Zheng1-GPRB14}. First, insight is gained by considering the effect of quantum fluctuations at high energy scales on low energy properties. 
An efficient way to study such effects is through a field-theoretic approach and bosonization. A 1D description is possible because for non-interacting leads the quantum dot couples to only an effectively 1D subset of the lead states. Introducing fermionic fields for these 1D electrons, we then proceed via phenomenological bosonization to describe the system in terms of bosonic fields (see Appendix\,\ref{sec:weak} for more detail). 
A natural way to view the result is that the dissipative environment mediates an interaction between the tunneling electrons. We wish to find the renormalization effects caused by this interaction; such effects are readily obtained for this model through the ``Coulomb-gas'' renormalization group (RG) \cite{KaneFisherPRB92,GoldsteinPRB10} technique.


We pause the main development at this point to make some brief comments about our approach. 
While bosonization tools have been extensively used to analyze electron transport, a number of potential pitfalls have been explored \cite{ShahBolechPRB16,BolechShahPRB16,FilipponeBrouwerPRB16,LeungDynamicalPRL95,BordaApplicabilityPRB08,SchillerAndreiX07,CulverAndreiX20-2} involving, e.g., properties far from a stable fixed point, high-energy cutoffs, or non-universal properties. In this work, we minimize these problematic issues by looking at universal properties close to the QCP, which is tuned to be stable.
In addition, while bosonization does not provide an exact description of fermionic quantum transport, there have recently been careful comparisons between numerical lattice and bosonized field-theoretic results for several simpler systems, for both equilibrium \cite{CamachoExactPRB19, LoCrossoverPRB19, AnthoreUniversalityRevEPJST20} and nonequilibrium \cite{KennesUniversalPRB14, BidzhievOutOfEqPRB19} properties. Furthermore, a detailed comparison of theory with experiment was carried out for the problem of tunneling through a single barrier (non-resonant tunneling, in contrast to our double-barrier resonant tunneling) in the presence of dissipation \cite{AnthorePierrePRX18}. 
The very good agreement found in all of these studies supports use of these techniques, which we apply here to study a system near its QCP. 


Returning to the main argument, one finds that the result of the RG analysis \cite{Mebrahtu12,Mebrahtu13,LiuRLdissipPRB14} is that when the dot is symmetrically coupled to the leads and is exactly on resonance, quantum fluctuations involving the dissipative environment \emph{enhance} transmission through the dot.
In a physically picturesque but somewhat loose sense, the barriers to transmission become effectively smaller as quantum fluctuations on more energy scales are taken into account.

The root cause is \emph{frustration} \cite{EggertAffleck92}. If the coupling is not symmetric ($t_S \!\neq\! t_D$), transmission is suppressed by quantum fluctuations involving the environment, as is normal for environmental effects on quantum tunneling. Effectively in an RG sense, the system is cut at the weaker link (larger barrier) and the dot is incorporated into the other lead [see Fig.\,\ref{structure}(c)]. In contrast, when the coupling is symmetric, frustration between incorporating the dot into the source versus the drain ensues. As a result, the dot becomes strongly hybridized with \emph{both} leads. This is the quantum critical state. 

The  properties of this critical state are heavily constrained because it is one of the two possible fixed points in the corresponding conformal field theory, namely the ``periodic'' fixed point \cite{WongAffleck94}. 
The constrained nature of the critical state leads to a broad universality: the $I$-$V$ curve near the QCP can be deduced from a variety of models that scale to the same critical state.  


\section{The quantum critical regime}
\label{sec:qc_regime}
 
For an explicit description of the strongly-hybridized limit that allows calculation of the $I$-$V$ curve, 
and as a result of the universality and enhanced transmission discussed in the last section, 
a wire with two weak potential barriers is a good model for the residual effect of the quantum dot \cite{KaneFisherPRB92,KaneFisherPRB92a,polyakov03}. Thus, consider two symmetric $\delta$-function barriers spaced by $\ell$ in a 1D wire of fermions described by the fields $\psi_R(x)$ and $\psi_L(x)$ for right and left movers. The fermions are then transformed via bosonization in the standard way into bosonic fields $\theta$ and $\phi$, obeying the  commutation relation 
$[\phi(x'),\partial_x\theta(x)]=i\pi\delta(x'-x)$ \cite{KaneFisherPRB92}. 
These fields represent the fluctuations in the density and phase of the fermions and for the moment are non-interacting. 
We add the effect of the environment and the bias below. 

The Hamiltonian, then, is $H_0 + H_T$ where 
\begin{eqnarray}
H_0 & = & \frac{1}{2} \int_{-\infty}^\infty dx \, 
[(\partial_x {\theta})^2 + (\partial_x {\phi})^2], 
\label{eq:H0_strong}\\
H_\textrm{\,T} & = & A \sum_\pm \cos[2\sqrt{\pi}\theta(\pm\ell/2) \pm k_F\ell] .
\label{eq:S_2}
\end{eqnarray}
The form  $\cos[2\sqrt{\pi}\theta]$ appears because it corresponds to $2k_F$ backscattering of the underlying fermions \cite{KaneFisherPRB92}, 
$\psi^\dagger_R\psi^{}_L + \text{h.c.}$, as can be checked by using the bosonization relation for $\psi_{L,R}$ [following the convention Eq.\,\eqref{eq:bosonization}]. Backscattering is the most important effect of scattering from a potential.
In writing Eqs.~\eqref{eq:H0_strong} and \eqref{eq:S_2}, we have chosen the bosonization convention near the strong-coupling fixed point (enhanced transmission implies weak barriers). In the following calculation, effects of $H_\textrm{\,T}$ are found to only leading order in $A$. In this way, the subtle concerns of bosonization consistency under different boundary conditions \cite{ShahBolechPRB16,BolechShahPRB16,FilipponeBrouwerPRB16} are avoided.

It is convenient to form the sum and difference fields
$\theta_{c} \equiv [\theta(\ell/2) + \theta(-\ell/2)]/2$ and 
$\theta_{f} \equiv [\theta(\ell/2) - \theta(-\ell/2)]/2$. 
When on resonance for a single level, one has $k_F\ell=\pi/2$ at strong coupling \cite{GiamarchiBook}, so that the barrier terms become  
\begin{equation}
H_\textrm{\,T}  =  A \cos(2 \sqrt{\pi} \theta_c ) \sin(2 \sqrt{\pi} \theta_f) .
\label{eq:STstrong}
\end{equation}

A key step in our argument is to incorporate the external bias potential $V$ and the fluctuating potential caused by the environmental field $\varphi$ in the strongly transmitting state. The environmental potential fluctuations are given by $\sqrt{2\pi}\dot{\varphi}$ which in a Hamiltonian formulation corresponds to 
$ir2\sqrt{2\pi}\partial_x\vartheta(0)$, where $\partial_x\vartheta(0)$ appears naturally as the charge fluctuation operator conjugate to $\varphi$ [see Eq.\,\eqref{eq:Henv}]. 
The effective fluctuating bias is, then, 
\begin{equation}
    e\tilde{V} \equiv eV +  ir2\sqrt{2\pi}\partial_x\vartheta(0).
\end{equation}
It is important to consider how this bias is expressed near the quantum dot. For large barriers (weak coupling) the 
potential difference drops between the source and drain leads, Eq.\,\eqref{eq:Hmu}. In contrast, near full transmission (strong coupling)  
the potential is applied between the \emph{right-moving} fermions (those coming from the source) and \emph{left-moving} fermions (from the drain). 
The fact that the potential drops in this way when a system is near full transmission is well-known, 
for instance, in discussing the quantum Hall effect in terms of edge states \cite{[{}][{, pp.\ 199, 310-4.}]{IhnBook}}. 
Thus, the bias and environmental coupling are
\begin{widetext} 
\begin{subequations}
\label{eq:Hbias-strong}
\begin{eqnarray}
H_{\mu+\text{Env}} & = &
\frac{e\tilde{V}}{2}
\left[  \int_{-\infty}^{-\ell/2} 
dx\, \psi^{\dagger}_R(x) \psi_R(x) - \int_{\ell/2}^{\infty}  
dx\, \psi^{\dagger}_L(x) \psi_L(x) \right] \quad \\
& =& \frac{e\tilde{V}}{4} \frac{1}{\sqrt{\pi}} \left[ \int_{-\infty}^{-\ell/2} dx [\partial_x\theta(x)-\partial_x \phi (x) ] 
- \int_{\ell/2}^{\infty} dx [\partial_x \theta (x) + \partial_x \phi (x)] \right] \\
& = & - \left[ \frac{eV }{2 \sqrt{\pi}} + 
r\sqrt{2}\,i\partial_x\vartheta(0) \right]
\int_{0}^{\infty} \!dx \;\partial_x \theta_c (x)  .
\label{bias_effect}
\end{eqnarray}
\end{subequations}
\end{widetext}

The dependence on external bias, the first term in Eq.\,\eqref{bias_effect}, is handled by performing a time-dependent gauge transformation that moves the bias into the barrier term $H_\textrm{\,T}$, as mentioned in connection with Eq.\,\eqref{eq:Hmu} [see also Eq.\,\eqref{eq:Hwt}]. Thus, in Eq.\,\eqref{eq:STstrong}
$\cos(2 \sqrt{\pi} \theta_c ) \to \cos(2 \sqrt{\pi} \theta_c +eVt)$.  Since the right-moving particles (from the source) have chemical potential $eV$ higher than that of the left-moving particles (from the drain), the bias appearing as a phase $eVt$ in the backscattering operator is quite natural.

The next step is to integrate out the environmental degrees of freedom $\varphi(x)$ and $\vartheta(x)$. The bilinear coupling to the lead fermions, the second term in Eq.\,\eqref{bias_effect}, then generates an effective coupling that causes $\theta_c(x)$ as well as its conjugate field denoted $\phi_f(x)$ to be \textit{interacting} fields. 
The integrating-out procedure is best performed in a Lagrangian formulation; it is straight forward using standard methods and given in Appendix\,\ref{sec:integrateout}. 
Finally, to emphasize that they are interacting fields, we relabel these fields $\theta_c'(x)$ and $\phi_f'(x)$. 

The effective description of the large transmission (strong-coupling) regime thus obtained is \cite{BiasLL}
\begin{subequations}
\begin{eqnarray}
H^\textrm{eff} &=& \frac{1}{2} \int_{0}^\infty dx
\Big[ (\partial_x {\theta}_f)^2 + (\partial_x {\phi}_c)^2 
\label{eq:Hboson-freenon}
\\
& &+ (1+r)(\partial_x {\theta}_c^\prime)^2
+ \frac{1}{1+r} (\partial_x {\phi}_f^\prime)^2 \Big]
\label{eq:Hboson-freeint}
\\[0.05in] 
&+& A \cos\left[2 \sqrt{\pi} \theta_c'(0) +eVt \right] \sin\left[2 \sqrt{\pi} \theta_f(0) \right]. \quad
\label{eq:Hboson-backscatter}
\end{eqnarray}
\label{eq:Hboson}
\end{subequations}
\hspace*{-0.095in}
We emphasize that the modes represented by fields $\theta_f$ and $\phi_c$ are \textit{free} while those represented by $\theta_c'$ and $\phi_f'$ are \textit{interacting}. 

The coupling between these two sets of modes is given by the barrier term, \eqref{eq:Hboson-backscatter}.
Recalling that a bosonic operator of the form $\cos (2\sqrt{\pi}\theta)$ corresponds to backscattering of the underlying fermions, we see that this coupling involves the  
\emph{simultaneous} backscattering of both sets of modes. 

The form of the barrier term \eqref{eq:Hboson-backscatter} is convenient for calculating the $I$-$V$ curve. Physically, it is also consistent with the form of the backscattering operator in resonant tunneling through a LL at zero bias, namely $\cos[2\sqrt{\pi} \theta'(0)] \partial_x \theta'(0)$ \cite{EggertAffleck92} where $\theta'(x)$ is the interacting field describing the LL. We show this explicitly in Appendix\,\ref{sec:LLbackscatter}. 
The link to LL physics is made by identifying the LL interaction parameter as $g\equiv 1/(1+r)$, the same expression \cite{SafiSaleurPRL04} as for tunneling through a single barrier in the presence of dissipation. 

The strength of the barrier term, $A$, is not known microscopically as it is the result of the RG flow from weak to strong coupling. It is helpful to recall at this point the equilibrium flow to the full transmission point. 
The equilibrium RG scaling equation for $A$ coming from Eq.\,\eqref{eq:Hboson-backscatter} can be derived by standard methods \cite{AltlandSimonsBook,GogolinBook}; it is 
$dA/d(\ln D) = A/(1+r)$,
where the energy cutoff $D$ runs from $D_0 = 1$ down to $0$.
The scaling dimension of the backscattering operator is then $\Omega \equiv 1 + 1/(1+r)$, showing that the operator is \emph{irrelevant} and $A \to 0$ at the QCP. Notice that the effect of the externally applied bias, $V$, is through this irrelevant operator. 

The linear response conductance at zero temperature is thus that of the system defined by only Eqs.\,\eqref{eq:Hboson-freenon}-\eqref{eq:Hboson-freeint}. By combining the charge and flavor fields, one obtains a system that is explicitly translationally invariant \cite{PerfectTransmission},  
in which one therefore has perfect transmission: $G=e^2/h$ \cite{Mebrahtu12}.

From general considerations (see e.g.\  \cite{EggertAffleck92,FisherGlazman97,AffleckLesHouches10}) one expects the deviation from perfect transmission at low temperature or bias to be a power law related to the scaling dimension of the leading irrelevant operator at the QCP. This is precisely the operator in Eq.\,\eqref{eq:Hboson-backscatter}. Using the scaling dimension $\Omega$ above, we expect 
$\left| dI/dV - e^2/h \right| \propto T^{2/(1+r)}$ or $\propto V^{2/(1+r)}$.

\section{The \emph{I-V} Curve} 
\label{sec:I-Vcurve}  

We now turn to an explicit calculation of the $I$-$V$ curve: we find the correction to perfect transmission caused by the joint backscattering term Eq.\,\eqref{eq:Hboson-backscatter} using a Keldysh nonequilibrium approach \cite{HaugJauho08chapter12}. Because $A$ is small, we work to leading order in this term but the bosonic fields are kept to all orders.  
This leads to a considerable simplification: a Keldysh calculation for scattering by a local operator to second order shows that the current is related to the backscattering rate $\Gamma(V, T)$ \cite{LevitovReznikovPRB04}. 


The backscattering matrix element needed is \cite{SassettiWeissEPL94}  
\begin{equation}
\begin{aligned}
\langle f | H_{\textrm{T}} | i \rangle = & 
A \, \langle R_1^f | \cos[2\sqrt{\pi}\theta^{\prime}_c(0)] | R_1^i \rangle 
\\ 
& \times \langle R^f_2 | \sin[2\sqrt{\pi}\theta_f(0)] | R^i_2 \rangle,
\label{matrix_element_extended}
\end{aligned}
\end{equation}
where $| R_1 \rangle$ and $| R_2 \rangle$ represent the states of $\theta_c^{\prime}$ and $\theta_f$, respectively, and $i$ and $f$ label the initial and final states. Recall that in time-dependent perturbation theory, an explicit oscillatory time dependence such as $eVt$ in \eqref{eq:Hboson-backscatter} factors out and enters the energy conservation constraint. The rate is, then, given by  
\begin{equation}
\begin{aligned}
\Gamma(V,T) =& A^2 \frac{2 \pi}{\hbar} \sum_{R^i_1 R^f_1} \sum_{R^i_2 R^f_2} \\
& \times | \langle R^{f}_1 | \cos\left[2\sqrt{\pi} \theta^{\prime}_c(0)  \right] | R^i_1 \rangle |^2 P_{\beta} (R^i_1)\\
& \times | \langle R^{f}_2 | \sin\left[2\sqrt{\pi} \theta_f (0)  \right] | R^i_2 \rangle |^2 P_{\beta} (R^i_2) \\ 
& \times \delta (E_{R^i_1}+E_{R^i_2}+e V-E_{R^f_1}-E_{R^f_2}), 
\end{aligned}
\label{gamma}
\end{equation}
where $P_{\beta}(R^i_{1,2}) = \langle R^i_{1,2} | \rho_{\beta} | R^i_{1,2} \rangle$ refers to the density matrices of the fields 
(in equilibrium) and the subscript $\beta$ is a reminder of the effect of temperature. The delta-function at the end of Eq.\,\eqref{gamma} guarantees energy conservation, where $E_{R_n^s}$ refers to the energy of the reservoir $R_n^s$ in the initial ($s = i$) or final ($s = f$) state. An explicit expression for this rate can be found by changing to the Heisenberg picture and using standard methods to evaluate the bosonic correlators, 
as outlined in Appendix~\ref{sec:BosonCor}.

The net current is related to the difference of this rate and that in the opposite sense, namely $\Gamma(-V, T)$. Since the energy associated with the bias in each backscattering event is $eV$, we see that the charge carried by each quasi-particle is $e$ \cite{FractionalCharge}.
Consequently, the backscattering-related current is $\Delta I(V,T) = e[\Gamma(V,T) - \Gamma(-V,T)]$. Adding this to the perfect transmission when $A=0$, we arrive at our final result for the $I$-$V$ curve,
\begin{equation}
\begin{aligned}
I (V,T)& = \frac{e^2}{h} V \left\{ 1 - \frac{ A^2 \pi^2}{\hbar^2 \omega_R^2} \frac{1}{\mathbf{\Gamma}(\frac{2}{1+r}+2)} \left(\frac{2\pi k_B T}{\hbar \omega_R}\right)^{\frac{2}{1+r}} \right. \\
& \qquad\qquad\quad \times \left. \left|
\frac{\mathbf{\Gamma} \left(\displaystyle{\frac{1}{1+r} + 1 +i \frac{e V}{2 \pi k_B T}}\right)}
{\mathbf{\Gamma}\left(\displaystyle{1 + i \frac{e V}{2 \pi k_B T}}\right)}\right|^2 \right\}, 
\end{aligned}
\label{eq:full_I-V}
\end{equation}
where $\mathbf{\Gamma}(x)$ is the Gamma function. 

\emph{Eq.\,\eqref{eq:full_I-V} is the main theoretical result of this paper:} the nonlinear $I$-$V$ curve to leading order in the backscattering amplitude $A$ in the critical regime of an 
interacting 
QCP (path 1 in Fig.\,\ref{structure}). 
Properties of this QCP are observable  
by tuning the system (described by the original microscopic Hamiltonian) to be on resonance and to have symmetric source and drain barriers. 
The renormalization caused by quantum fluctuations of the dissipative environment (the flow to the QCP) is cutoff by the temperature. The applied bias $V$ appearing in the irrelevant operator Eq.\,\eqref{eq:Hboson-backscatter} likewise limits the approach to the QCP. 
At large bias, Eq.\,\eqref{eq:full_I-V} yields a power-law dependence, $\left| dI/dV -e^2/h \right| \propto V^{2/(1+r)}$, as expected from the equilibrium RG analysis above. 
A plot of the full result is shown in Fig.\,\ref{fig:GV}.

\bigskip
\section{Comparison to experiment}
\label{sec:experiment}

\begin{figure}
\begin{center}
\vspace*{-0.2cm}
\includegraphics[width=8.6cm]{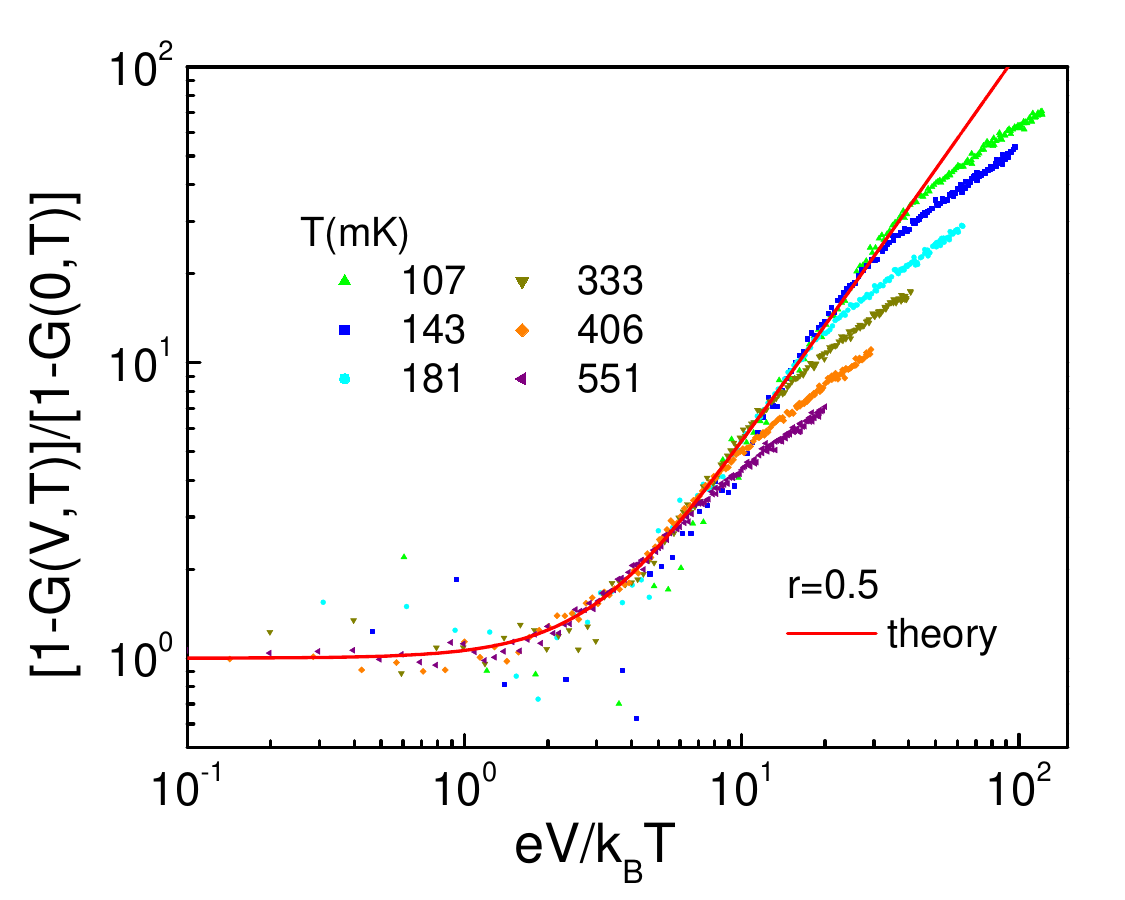}
\end{center}
\par
 \vskip -0.7cm
\caption{
Conductance 
measured in the full bias range---from much smaller to much larger than the temperature---presented as the deviation from perfect conductance $1\!-\!G(V,T)$ scaled by $1\!-\!G(V\!=\!0,T)$ and plotted vs.\ the rescaled bias $eV/k_B T$. Here $G(V,T)$ is the differential conductance $G\!=\!(h/e^2)\,dI/dV$ and $r \!=\!0.5$. 
Gate voltages are tuned to their critical values; thus, for small $V$ and $T$ the system approaches its QCP (see path 1 in Fig.\,\ref{structure}).
The symbols are the experimental results at the color-coded temperatures. The red line is the result of the non-equilibrium theory [Eq.\,\eqref{eq:full_I-V}], in which there are no free parameters. Note the excellent agreement between the theory and data in both the crossover and power-law regimes. At larger $V/T$, non-universal effects begin to set in due to $1\!-\!G(V,T)$ becoming non-negligible compared to $1$. 
}
\label{fig:GV}
\end{figure}

Experiments were performed on quantum dots fabricated from carbon nanotubes contacted by Cr/Au electrodes. The electrodes were further connected to the bonding pads by Cr resistors that provided dissipation. For more information on the fabrication and characteristics of the samples, see Refs.\,\cite{Mebrahtu12,Mebrahtu13}. Here we show data from a sample with $r\!=\!0.5$ (for similar data for a sample with $r\!=\!0.75$ see 
the supplemental material  \cite{SupMat}). 
The value of $r$ is determined in an independent equilibrium measurement of $G(T)$ off resonance; 
in this case, the (co-)tunneling effectively reduces to the single-barrier case and $G \propto T^{2r}$  \cite{Mebrahtu12}. Once $r$ is fixed, we check that the equilibrium ($eV \!\ll\! k_BT$) value of $1-\!G$ on resonance scales as $T^{2/(1+r)}$, as demonstrated previously \cite{Mebrahtu13}. 
This confirms that the gate voltages controlling the level's energy and the symmetry of the barriers are tuned to their critical values. 

For these critical values,
we consider the conductance in the full range of applied bias---both smaller and larger than $k_B T$, corresponding to the equilibrium and non-equilibrium regimes, respectively. Fig.\,\ref{fig:GV} shows the conductance $G$, measured in units of $e^2/h$ and rescaled such that at a given temperature $1-G(V)$ is divided by $1-G(V\!=\!0)$. The experimental results are compared to the theoretical result Eq.\,\eqref{eq:full_I-V} (solid line).  
(For other ways of plotting the data, including without rescaling, see Appendix~\ref{sec:expt-extra} and \cite{SupMat}.)
Notably, $A$ and $\omega_R$ in \eqref{eq:full_I-V} have been eliminated by taking the ratio $[1-G(V)]/[1\!-\!G(V\!=\!0)]$. We see that the theory curve captures the experimental behavior remarkably well without any fitting parameters.

Comparing closely the experimental and theoretical results, we see two striking features of the theory: first, it captures the \emph{crossover regime} $eV \!\!\sim\! kT$ very accurately, and, second, it yields the correct \emph{prefactor} of the universal $\propto V^{2/(1+r)}$ dependence at high bias.
Thus our theory goes well beyond the frequently used scaling arguments that produce only the exponent in the scaling regime (the slope on this log-log plot). 
(As a contrast, a naive scaling argument is sketched in Appendix\,\ref{sec:RGforIV}.)
\emph{The excellent agreement between theory and experiment in a wide range of $eV/k_BT$ is a striking confirmation of our far-from-equilibrium calculation.}

It is important to realize that, unlike measurements that use a weakly coupled electrode as a tunnel probe that measures the equilibrium density of states at finite bias (for example \cite{LeturcqEnsslinPRL05,Makarovski07b,Potok2CK07,ChangMarcusPRL13,KellerDGGNat15}), here the two biased leads remain equally coupled to the quantum dot, creating genuinely nonequilibrium conditions \cite{MeirWingreenPRL92,PustilnikPRB04}. 

At high enough $eV/k_BT$, the experimental curves deviate from the theoretical prediction (Fig.\,\ref{fig:GV}). There are several possible contributions to this deviation. Because $1\!-\!G$ is no longer small, irrelevant operators not included in our effective 
model of the QCP  
[Eq.\,\eqref{eq:Hboson}] may become significant. An additional possible contribution is that our second-order analysis is inadequate at high bias. At the same time, note that the range of applicability of our analytical results is pushed to higher and higher $eV/ k_BT$ as the temperature is lowered. 


\section{Non-Fermi-liquid state: Tunneling in an Environment}
\label{sec:DCB}

To enhance physical understanding of our main results, 
we rewrite our description of the quantum critical regime, Eq.\,\eqref{eq:Hboson}, in terms of non-interacting fermions coupled to an environment. 
An outline of the main features of the argument is given here with the details provided in Appendix\,\ref{sec:derriveDCB}.  

In order to arrive at non-interacting fermions, we refermionize the non-interacting bosonic fields $(\theta_f,\phi_c)$  by simply using the bosonization relation in reverse. The free part of the bosonic Hamiltonian \eqref{eq:Hboson-freenon} maps to free right and left moving fermions on a half-infinite 1D wire. The factor $\sin\left[2 \sqrt{\pi} \theta_f(0) \right]$ in the interaction \eqref{eq:Hboson-backscatter} represents backscattering between the right and left fermions at the end of the wire (with a phase shift). 
Writing the factor $\cos\left[2 \sqrt{\pi} \theta_c'(0) + eVt \right]$ as two exponentials, we see that to leading order, \eqref{eq:Hboson-backscatter} can be viewed as an interaction with an environment given by the interacting bosonic fields $(\theta_c',\phi_f')$ together with a bias applied to the fermions. 
In the scattering process, the fermion can gain or lose energy $eV$ together with the corresponding excitation of the bosonic field.

This is a surprisingly simple view of this non-equilibrium non-Fermi-liquid! One of the first concrete examples of a non-Fermi-liquid was, of course, an electron tunneling in the presence of an environment \cite{WeissBook}---the single barrier version of our starting model. Here we see that the physics of the high transmission quantum critical state---one that involves strong renormalization effects---can also be viewed as a particle scattering at a point coupled with an environment. It is highly non-trivial that all the messy relaxation and dephasing associated with exciting a non-Fermi-liquid can be neatly summarized in such a compact and suggestive fashion.

More concretely, we can find the $I$-$V$ curve from this point of view: tunneling of non-interacting particles through a barrier in the presence of an environment \cite{IngoldNazarov92, NazarovBlanterBook, DevoretEsteveUrbina95}, albeit with a strange barrier and strange environment. The tunneling through the barrier consists of backscattering between two chiral fermion modes, and the environment $\theta_c'$ involves a nonlinear combination of the original electrons and environment. Nevertheless, the standard techniques of dynamical Coulomb blockade theory \cite{IngoldNazarov92, NazarovBlanterBook, DevoretEsteveUrbina95} can be applied to obtain the nonlinear $I$-$V$ curve to second order in $A$. 
We perform this calculation in Appendix\,\ref{sec:derriveDCB} and demonstrate that the result \cite{SassettiWeissEPL94, ZhengSScom98} is identical to that found from the bosonic description, Eq.\,\eqref{eq:full_I-V}. 

\section{Conclusions}
\label{sec:conclusions}

We have carried out an analytic calculation of a far-from-equilibrium $I$-$V$ curve  
for a system whose control parameters are tuned to 
a QCP (path 1 in Fig.\,\ref{structure}), and then presented experimental results enabling a detailed theory-experiment comparison. 
The calculation proceeds via an effective bosonic description 
valid near the full-transmission (strong-coupling) QCP.
The comparison of the resulting $I$-$V$ curve with the experiment validates this approach. Indeed,  
as shown in Fig.\,\ref{fig:GV}, the agreement with the experimental results throughout the crossover and asymptotic regimes is excellent. 

A simple physical interpretation is possible because only one of the charge modes in the system couples to the resistive environment, leaving the mode corresponding to fluctuations of the total charge in the dot free. 
This feature is not present, for instance, in the related problem of resonant tunneling in a Luttinger liquid. It allows us to find the $I$-$V$ curve, alternatively, from the problem of tunneling between left- and right-moving non-interacting fermions in the presence of a modified environment. 

To our knowledge, this is the first 
direct comparison of theory and experiment for 
a nonequilibrium $I$-$V$ curve of a system tuned to 
an interacting quantum critical point. 
A remarkable aspect of this system is that it is fully accessible to both 
calculation and measurement, 
allowing for a detailed comparison. 
This accessibility is characteristic of other nanoscale systems exhibiting boundary QPT, see e.g.\ \cite{KellerDGGNat15,IftikharPIerreNat15,IftikharPierreScience18,AnthorePierrePRX18, AnthoreUniversalityRevEPJST20, LeHurDrivenQImpReviewCRP18, MocaZarandPRL19, ChoiSim4DotsPRB20}, 
one of the reasons for increasing interest in this topic. 
As nonequilibrium results in quantum critical states are exceedingly rare, our results provide a valuable bench mark and test case for future studies of nonequilibrium steady states.

\begin{acknowledgements}
We thank E. Novais for helpful discussions.
The work in Taiwan (CHC, CYL) was supported by the NSC grants
No.98-2918-I-009-06 and No.98-2112-M-009-010-MY3, the MOE-ATU program, and the NCTS of Taiwan, R.O.C. 
The work in the U.S.A.\ was supported by the 
U.S.\ Department of Energy, Office of Science, Office of Basic Energy Sciences, Materials Sciences and Engineering Division, 
under Awards Nos.\ DE-SC0005237 (GZ, HUB), DE-SC0002765 (CTK, HM, GF), and DE-FG02-02ER15354 (AIS). 
\end{acknowledgements}

\appendix

\section{Effective Interactions at Weak Tunneling \\Through Bosonization}
\label{sec:weak}

In Sec.\,\ref{sec:qc_regime}, we show that by integrating out the environmental field $\vartheta$, the system at strong hybridization can be described by the effective Hamiltonian \eqref{eq:Hboson}, where a pair of canonical fields ($\theta_c'$ and $\phi_f'$) become interacting. This pair is similar to fields in a LL. 
In this appendix, we 
highlight that the fact that these fields are \textit{interacting} can also be understood in the weak tunneling regime \cite{Mebrahtu12,Mebrahtu13,LiuRLdissipPRB14}---the starting point of any perturbative RG analysis as it corresponds to weak coupling.

In the weak tunneling regime, bosonization is possible because an impurity couples to only an effectively 1D subset of lead states (for non-interacting electrons). We label these semi-infinite 1D leads $x\!\in\!(-\infty,0)$ for the source lead (S) and $x\!\in\!(0,+\infty)$ for the drain (D). 
These leads are modeled using 1D fermionic fields $c^{\dagger}_{\alpha, L/R}(x,t)$ with Fermi velocity set equal to one, where $\alpha=S/D$ labels the leads and $L/R$ indicates left- or right-moving particles. 

We proceed via phenomenological bosonization in the standard way \cite{GiamarchiBook,KaneFisherPRB92}, choosing the conventions of Ref.\,\cite{KaneFisherPRB92}:
\begin{equation}
c^{\dagger}_{\alpha, L/R}(x,t) = 
e^{\pm {\it i}k_F x} \frac{F_{\alpha}}{\sqrt{2\pi a_0}} 
e^{{\it i}\sqrt{\pi}[{\phi_\alpha}(x,t) \pm
{\theta}_\alpha(x,t)]} 
\label{eq:bosonization}
\end{equation}
where $\pm$ in the exponent corresponds to $L/R$.
$\phi_{\alpha}$ and $\theta_\alpha$ are conjugate
bosonic operators that describe electronic states in
the semi-infinite leads, obeying the standard commutation relation 
$[\phi(x'),\partial_x\theta(x)]=i\pi\delta(x'-x)$.
$a_{0}$ is a 
regularization scale for short distance or time, 
and the $F_{\alpha}$ are Klein factors. 
These latter can be simply carried along in the present problem, giving rise to no additional phases, and so we do not discuss them further. 
In bosonic form, the electron density is 
\begin{equation}
    \rho_{L/R}(x) = [ \pm \partial_x \phi(x) + \partial_x \theta(x) 
+ k_F/\sqrt{\pi}]/(2 \sqrt{\pi}).
\end{equation}
It is convenient to form the charge and flavor fields  \cite{FendleyPRB95},
\begin{equation}
\begin{aligned}
\phi_{f/c}(x) & \equiv \tfrac{1}{2} \left[ \phi_S(-x) \mp \phi_D(x) \pm \theta_S(-x) -\theta_D(x) \right] \\
\theta_{c/f}(x) & \equiv \tfrac{1}{2} \left[ \pm \phi_S(-x) + \phi_D(x) + \theta_S(-x) \pm \theta_D(x) \right] .
\end{aligned}
\label{eq:1d_new_fields}
\end{equation}
Note that $\phi_f(x)$ is conjugate to $\theta_c(x)$ and likewise $\phi_c(x)$ to $\theta_f(x)$, as in the strong-hybridization limit.


The tunnel Hamiltonian, Eq.\,\eqref{eq:HT}, is the key term in which to use the bosonization relation. Since the QCP occurs at symmetric coupling, we take identical coupling to the source and drain leads, $t_{S}\!=\!t_{D}\equiv t$. With symmetric tunneling and capacitance, the bias voltage drops symmetrically as well. Remembering the time-dependent gauge transformation from the end of Sec.\,\ref{sec:model}, 
we find that the tunneling term becomes 

\begin{equation}
\begin{aligned}
H_\textrm{T+$\mu$} & =\frac{t}{\sqrt{2 \pi a_0}} \left[ F_S d e^{i \sqrt{\pi} \phi_c} e^{i \left(\sqrt{\pi} \phi_f -\sqrt{\frac{\pi}{2}} \varphi +eV t/2 \right)} \right. \\
&  + \left. F_D d e^{i \sqrt{\pi} \phi_c} e^{-i \left(\sqrt{\pi} \phi_f -\sqrt{\frac{\pi}{2}} \varphi +eV t/2 \right)} + \textrm{h.c.} \right].
\end{aligned}
\label{bos_wt}
\end{equation}
All fields in Eq.\,\eqref{bos_wt} are taken at $x\!=\!0$, and we have used $\theta_{S/D}(0)\!=\!0$ \cite{KaneFisherPRB92,HonerWeiss2010} due to the Dirichlet boundary condition at the $x\!=\!0$ end of the leads because of the large barrier ($t$ is small due to weak coupling). 

Notice that the fields $\phi_f(x\!=\!0)$ and $\varphi$ enter in the same way in Eq.\,\eqref{bos_wt}, so it is natural to combine them via the transformation 
\begin{equation}
\begin{aligned}
\phi_f^{\prime}(x)&= \phi_f(x)-\frac{1}{\sqrt{2}}\varphi(x)\\
\varphi^{\prime}(x)&=\sqrt{r} \phi_f(x)+\frac{1}{\sqrt{2r}} \varphi(x).\\ 
\end{aligned}
\end{equation} 
Since the field $\varphi'$ completely decouples from the problem, we drop it from further consideration. After carrying out all these transformations on the free part of the Hamiltonian as well, the final expression for the Hamiltonian at weak tunneling is 
\begin{widetext}
\begin{equation}
\begin{aligned}
H_\textrm{Dot} & +  H_{\textrm{Leads+Env}}^{\textrm{eff}} = 
\epsilon_\textrm{d} d^\dagger d 
 +  \frac{1}{2}  \int_{0}^\infty  dx
\Big[ (\partial_x {\theta}_f)^2 + (\partial_x {\phi}_c)^2 
+ (1+r)(\partial_x {\theta}_c^\prime)^2
+ \frac{1}{1+r} (\partial_x {\phi}_f^\prime)^2 \Big] 
\\
H_\textrm{\,T+$\mu$} & = \frac{t}{\sqrt{2 \pi a_0}} \Big\{ F_S d e^{i \sqrt{\pi} \phi_c} e^{i (\sqrt{\pi} \phi^{\prime}_f +eV t/2)}  + F_D d e^{i \sqrt{\pi} \phi_c} e^{-i (\sqrt{\pi} \phi^{\prime}_f +eV t/2)} +\textrm{h.c.} \Big\}.
\end{aligned}
\label{eq:Hwt}
\end{equation}
\end{widetext}
Thus we see that the coupling of each tunneling electron to the environment generates an effective interaction between them. As in the quantum critical regime treated in Sec.\,\ref{sec:qc_regime}, \emph{one} of the sets of lead fields, $(\phi_f',\theta_c')$, becomes interacting. In contrast, in a LL, \emph{both} sets of lead fields would be interacting, an important distinction for interpreting our results in terms of the dynamical Coulomb blockade (see Sec.\,\ref{sec:DCB} and Appendix\,\ref{sec:derriveDCB}). 


When the dot is symmetrically coupled to the leads and is exactly on resonance ($\epsilon_\textrm{d}\!=\!0$), this weak-coupling description renormalizes to a strong-coupling fixed point that marks the QCP \cite{Mebrahtu12}. A ``Coulomb-gas'' RG \cite{KaneFisherPRB92,GoldsteinPRB10} treatment shows that $t$ becomes larger, suggesting that the tunnel barrier disappears and the system is effectively becoming increasingly uniform.
This occurs because of the presence of frustration
(see Sec.\,\ref{sec:qflucts} for discussion).
The strong coupling point is described by the effective Hamiltonian Eq.\,\eqref{eq:Hboson} \cite{ImpurityEntropy}.  
Note that it contains the same pair of interacting fields $(\phi_f',\theta_c')$, thus making the connection between the weak tunneling and strongly hybridized limits.

In contrast, with any asymmetry present, the fixed point to which the system flows corresponds to cutting the system at the weaker link and incorporating the dot into the other lead [see Fig.\,\ref{structure}(c)].
In this limit, the system can flow from the non-Fermi liquid strong coupling fixed point studied here to a Fermi liquid fixed point as one 
decreases 
the temperature or source-drain voltage compared to the energy scale of the asymmetry \cite{GuEduHarInprep20}.
Such a crossover to Fermi-liquid ground states has been analyzed previously in two impurity and two channel Kondo systems \cite{SelaExactTransPRL09, SelaNoneqQdotsPRB09, MitchellSelaPRL16}.


\section{Integrating Out the Dissipation Near \\Full Transmission}
\label{sec:integrateout}

We briefly show the final step in arriving at the effective strong-hybridization model Eq.\,\eqref{eq:Hboson} by integrating out the environment. We begin by rewriting the free lead and environment Hamiltonian, Eqs.\,\eqref{eq:Henv} and \eqref{eq:H0_strong}, in the form of an action,
\begin{equation}
\begin{aligned}
& S_0  = S_{\text{Leads}} + S_{\text{Env}} \\
& = \frac{1}{2} \iint d\tau dx [ (\partial_x \theta_c)^2 + (\partial_{\tau} \theta_c)^2 + (\partial_x \theta_f)^2 + (\partial_{\tau} \theta_f)^2 ] \\
&\quad + \frac{1}{2} \iint d \tau dx\; 2r\,[ (\partial_x \vartheta)^2 + (\partial_{\tau} \vartheta)^2 ].
\end{aligned}
\label{eq:free_action}
\end{equation} 
Here we write the action in terms of the $\theta$ fields rather than $\phi$ because of the boundary conditions connected to the very weak barrier (strong hybridization) \cite{KaneFisherPRB92}.
Since the action is quadratic except for the backscattering occurring at the origin, the $x\!\neq\!0$ degrees of freedom can be integrated out. 
If the backscattering is not too strong, the free action is minimized when $\theta(x,\omega_n) = \theta(x\!=\!0,\omega_n) \exp(-|\omega_n x |)$ and $\vartheta(x,\omega_n) = \vartheta(x\!=\!0,\omega_n) \exp(-|\omega_n x |)$, where $\omega_n$ is the Matsubara frequency \cite{KaneFisherPRB92}. We can thus integrate out the $x\neq0$ part of the system so that the free action becomes zero-dimensional,
\begin{equation}
S_0 = \frac{1}{\beta}\sum_{\omega_n} |\omega_n|[\theta_f(\omega_n)^2 + \theta_c(\omega_n)^2 + 2 r \vartheta(\omega_n)^2],
\label{eq:free_action_euclidean}
\end{equation}
where all the fields are evaluated at $x=0$ \cite{QBM}.

The coupling between the leads and the environment is given by the second term in Eq.\,\eqref{bias_effect}, which corresponds to
\begin{equation}
\begin{aligned}
S_{\text{coup.}} & = 
- i 2\sqrt{2}\int d\tau\,   r\, [\partial_x\vartheta(0)]\, \theta_c(0)\\
& =  i \frac{2\sqrt{2}\, r}{\beta}\sum_{\omega_n} |\omega_n| \vartheta ({\omega_n}) \theta_c(-\omega_n),
\end{aligned}
\label{eq:action_with_dissipation}
\end{equation}
rewritten in Matsubara summation form. Since this term is a quadratic product of $\vartheta$ and $\theta_c$, we can easily integrate out the environment $\vartheta$ with a Gaussian path integral.
The integral is done with the partition function,
\begin{equation}
Z = \iiint D[\theta_c] D[\theta_f] D[\vartheta]  
e^{-S_0[\theta_c, \theta_f, \vartheta] 
- S_{\text{coup.}} [ \theta_c, \vartheta ] 
- S_{\text{T}} [ \theta_c, \theta_f ]}, 
\label{eq:partition_function}
\end{equation}
where $S_\text{T}$ is the contribution of the tunneling term, Eq.\,\eqref{eq:STstrong}, to the action. 
The effective partition function thus becomes
\begin{equation}
Z^{\text{eff}} = \iint D[\theta_c] D[\theta_f] e^{-S'_{\text{Leads}} - S_{\text{T}}},
\label{eq:effective_partition_function}
\end{equation}
\vspace*{-0.3cm}
where 
\begin{equation}
\begin{aligned}
S'_{\text{Leads}} &= \frac{1}{2} \iint d\tau dx 
\left[ (\partial_x \theta_f)^2 + (\partial_{\tau} \theta_f)^2 \right.\\
 &\left. \quad + (1+r) (\partial_x \theta_c)^2 + (1+r) (\partial_{\tau} \theta_c)^2 \right] .\\
\end{aligned}
\label{eq:effective_free_action}
\end{equation}
Here we have extended the fields back to their original semi-infinite domains. Notice that the interaction between the dissipative environment and the $\theta_c$ field has been effectively incorporated into the free action of $\theta_c $ so that it becomes effectively interacting with strength $1/(1 + r)$. Finally, we convert to the Hamiltonian form and, to be consistent with the notation of the main text, relabel $(\theta_c, \phi_f )$ as $(\theta_c', \phi_f' )$.

Strictly speaking, when the system is non-equilibrium, the effective Hamiltonian Eq.\,\eqref{eq:effective_free_action} is incomplete: we have ignored the applied bias term [the first term of Eq.\,\eqref{eq:Hboson-backscatter}] which is also linear with respect to $\theta_c$.
When we include its effect, the bias is modified to an effective value $V/(1+r)$, similar to the case in a Luttinger liquid wire with interaction $g = 1/(1+r)$.
This 
factor $1/(1+r)$, however, \textit{disappears} when we take into consideration the wire-reservoir boundary condition \cite{PonomarenkoRPB95,SafiSchulzPRB95,MaslovStonePRB95,FrohlichPRB96}.
After realizing the above fact,
the relabeling should also be carried out, of course, in the backscattering term, 
Eq.\,\eqref{eq:STstrong},  
$H_\textrm{\,T}  \to  A \cos(2 \sqrt{\pi} \theta_c' +eVt ) \sin(2 \sqrt{\pi} \theta_f)$. 
We thus arrive at Eq.\,\eqref{eq:Hboson}.

\section{Connection to Luttinger Liquid Backscattering Operator}
\label{sec:LLbackscatter}

\vspace*{-0.1cm}
As mentioned in the introduction, the problem of resonant tunneling in a LL is similar in many respects to the problem of dissipative resonant tunneling that we study here. In a LL, the barrier term that arises in the strong hybridization regime---the term analogous to \eqref{eq:Hboson-backscatter} in $H^\textrm{eff}$---is $\cos[2\sqrt{\pi} \theta'(0)] \partial_x \theta'(0)$, where $\theta'(x)$ is the interacting field describing the LL \cite{EggertAffleck92}. Note that in a LL both the $c$ and $f$ modes are interacting and related to $\theta'(x)$. The connection between the LL 
barrier term above and 
the operator in our dissipative system
can be obtained explicitly as follows.  

To arrive at the LL expression from Eq.\,\eqref{eq:Hboson-backscatter}, we expand about the midpoint of the two barriers and call this point $x\!=\!0$. Then with the definition of the common and difference fields, we have $\theta'_c \approx \theta'(0)$ and $\theta'_f \approx \partial_x \theta'(0) \ell/2$.
Since $\partial_x \theta'(0) \ell/2$ is small and fluctuating, the $\sin\left[2 \sqrt{\pi} \theta_f(0) \right]$ factor is simply expanded to yield $\pi^{3/2} \partial_x \theta'(0)/2k_F$, where we have used the resonant requirement $k_F\ell\!=\!\pi/2$. 
Combining this with $\cos\theta'_c \approx \cos\theta'(0)$, we find the expression above for the near-resonance backscattering from two barriers in a LL. Thus, the operator that we use 
and the operator in the LL case are physically consistent.

\begin{figure*}[t]
    \par
 \vskip -0.5cm
  \centering
    \includegraphics[width=0.85\textwidth]{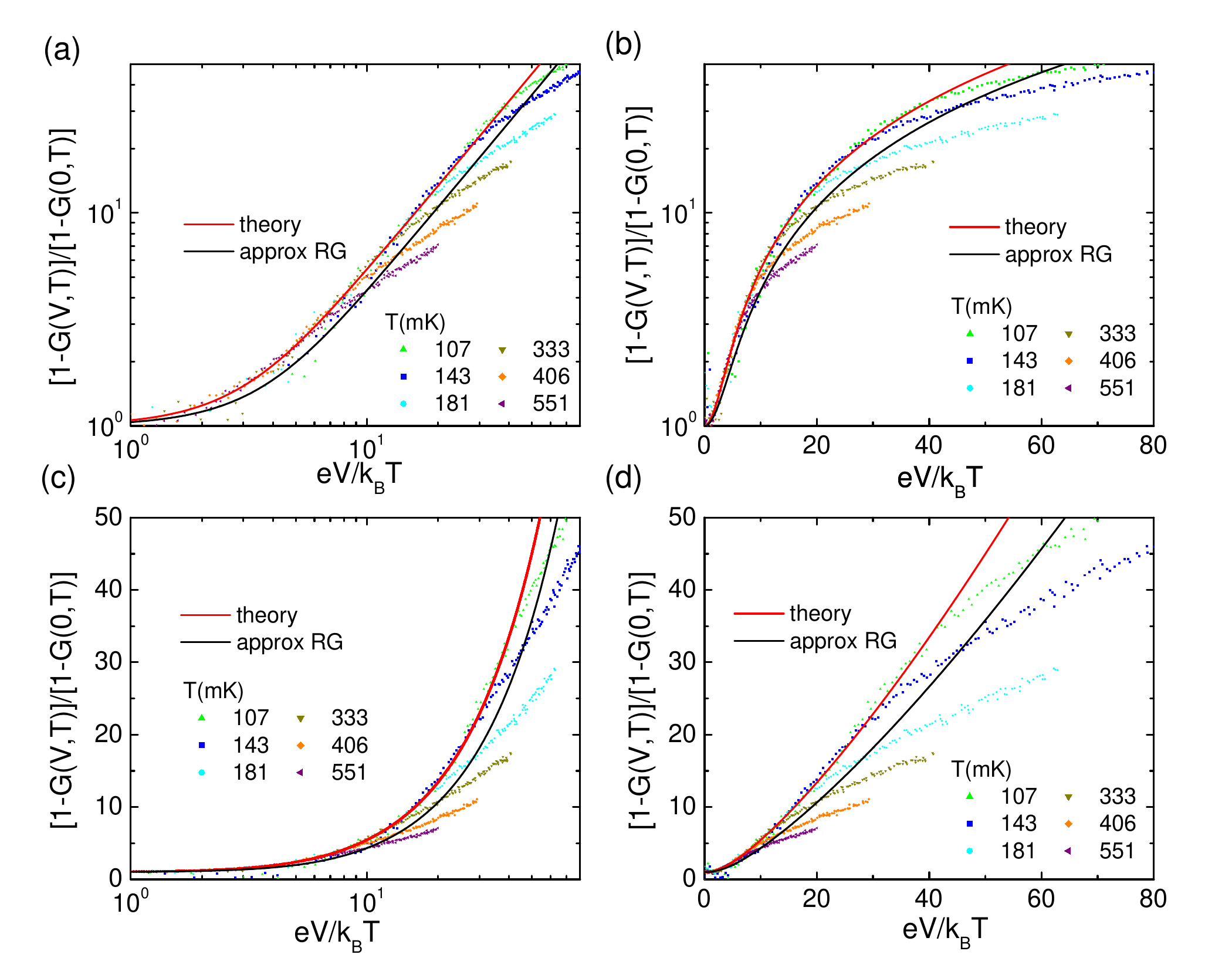}
    \par
 \vskip -0.5cm
  \caption{Comparison between the experimental data and theoretical calculations with dissipation $r \!=\! 0.5$ (same data as in in Fig.\,\ref{fig:GV}). Our theoretical result, Eq.\,\eqref{eq:full_I-V}, is the red line; the approximate RG treatment of Appendix\,\ref{sec:RGforIV} is shown in black. We emphasize the excellent agreement of the theoretical curve (red line) with the experimental data in the crossover regime.}
  \label{fig:S1}
\end{figure*}

\vspace{0.1cm}
\section{Evaluation of Bosonic Correlators}
\label{sec:BosonCor}

In this section, we give an explicit expression for the tunneling rate Eq.\,\eqref{gamma}. First, 
we introduce the $\delta$-function identity
\begin{equation}
\begin{aligned}
 & \delta (E^i+E_R^i+e V-E^{f}-E_R^f)\\
& =  \frac{1}{2\pi\hbar} \int_{-\infty}^{\infty} \!\!dt \exp\left[\frac{i}{\hbar} (E^i+E_R^i+e V-E^{f}-E_R^f) t\right]. 
\label{delta_fourier}
\end{aligned}
\end{equation} 
Notice that the factors $\exp(iE_{R_n^s} t/\hbar)$ can be produced by acting on the initial ($s \!=\! i$) or final ($s \!=\! f$) state with $\exp(iHt/\hbar)$. 
Changing to the Heisenberg picture for the fields, we thus find
\begin{widetext}
\begin{equation}
\begin{aligned}
\Gamma(V,T) &\! =\! \frac{A^2}{\hbar^2} \int^{\infty}_{-\infty} \!\!dt \sum_{R^i_1 R_1^{f}} \langle R^i_1 | \cos\left[2\sqrt{\pi} \theta^{\prime}_c(t) \right] | R^{f}_1 \rangle  \langle R^{f}_1 | \cos\left[2\sqrt{\pi} \theta^{\prime}_c(t=0) \right] | R^i_1 \rangle P_{\beta}(R^i_1) \\
& \qquad \times \sum_{R^i_2 R_2^{f}} \langle R^i_2 | \sin\left[2\sqrt{\pi} \theta_f(t) \right] | R^{f}_2 \rangle  \times\langle R^{f}_2 | \sin\left[2\sqrt{\pi} \theta_f(t=0) \right] | R^i_2 \rangle P_{\beta}(R^i_2) e^{i e V t/\hbar}\\[0.2cm]
& \!= \! \frac{A^2}{\hbar^2}\! \int^{\infty}_{-\infty} \!\!dt\, e^{i e V t/\hbar} \!\!
\left\langle \cos\left[2\sqrt{\pi} \theta_c^{\prime} (t) \right] 
\cos\left[2\sqrt{\pi} \theta_c^{\prime} (0) \right] \right\rangle \!
\left\langle \sin\left[2\sqrt{\pi} \theta_f (t) \right] 
\sin\left[2\sqrt{\pi} \theta_f (0) \right] \right\rangle , 
\end{aligned}
\label{gamma2}
\end{equation}
where we have dropped the argument $x \!=\! 0$ for clarity. 
Evaluation of the bosonic correlation function is standard, see for example Refs.\,\cite{GiamarchiBook,IngoldNazarov92}. In terms of the scaling dimension $\Omega \!=\! 1 + 1/(1+r)$ of the backscattering operator,  
the result for the rate is
\begin{equation}
\begin{aligned}
\Gamma(V,T)&\!=\!\frac{A^2}{4 \hbar^2} \int^{\infty}_{-\infty} \!\!dt\, 
\!e^{i e V t/\hbar} 
\exp\!\!\left[ -2\Omega \ln \sinh \left(\frac{\pi k_B T |t|}{\hbar} \right) \right.\!+ \! 2\Omega \ln \frac{\pi k_B T}{\hbar \omega_R} 
\! -\! \Omega i \pi \text{Sign} (t) \! - \! 2\Omega \gamma \bigg]  \\[0.2cm]
&= \frac{A^2}{4 \hbar^2} \frac{\pi}{\Gamma(2\Omega)} \left(\frac{2 \pi k_B T}{\hbar \omega_R}\right)^{2\Omega-1} \frac{1}{\omega_R}  \exp\left(\frac{e V}{2 k_B T}\right) \bigg|\mathbf{\Gamma}\left(\Omega + i \frac{e V}{2 \pi k_B T}\right)\bigg|^2,
\end{aligned}
\label{gamma3}
\end{equation}
\end{widetext}
where $\omega_R$ is the energy cutoff of the bosonic bath, $\gamma$ is Euler's constant, and $\mathbf{\Gamma}(x)$ is the Gamma function. Physically, as this rate involves gain of energy, it corresponds to backscattering from the right-moving to left-moving channel [using the convention of Eqs.\,\eqref{eq:Hmu} and \eqref{eq:Hbias-strong}].

\section{Experimental Data Replotted} 
\label{sec:expt-extra}
\vspace*{-0.1cm}

To supplement the comparison between experimental data and the theoretical results, we provide plots of the data from Fig.\,\ref{fig:GV} using different combinations of log and linear scales. 
For the $r=0.5$ case, in Fig.\,\ref{fig:S1} we plot in four different ways the deviation of the differential conductance from perfect $e^2/h$: $[1-G(V,T)]/$ $[1-G(0,T)]$ vs.\ $eV/k_BT$ is plotted on 
(a) log-log, (b) semi-log, (c) linear-log, and (d) linear-linear scales. 
Focusing on the crossover regime, note the excellent agreement between the experimental data and the full theoretical results (red line).

\section{Approximate RG Argument for \textit{I-V} Curve}
\label{sec:RGforIV}
\vspace*{-0.1cm}

It is interesting to compare the data and full theoretical results to  
a much simpler but approximate treatment of the $I$-$V$ curve that can be developed starting from the equilibrium RG equation, $dA/d(\ln D) = A/(1+r)$.
Note that $A$ is thus energy ($\epsilon$) dependent, 
$A (\epsilon) = A_0 \epsilon^{1/(1+r)}$ where $A_0$ is a constant. This power-law scaling is cut off below $T$, making $A$ temperature dependent as well, $A(\epsilon,T)$. The differential conductance $G(V,T)\!=\!dI/dV$ can be obtained approximately by integrating the spectral function of the transmission probability $T(\epsilon,T) = 1 - R(\epsilon,T)$ over $\epsilon$ 
with $R\propto {A}^2$. A more accurate but technically much  more complex RG treatment would involve computing $R(\epsilon,T,V)$ out of equilibrium at a finite bias $V$. This has been done for the single-channel Kondo model 
\cite{RoschNoneqKondoPRL03,RoschNoneqKondoJPSJ05} and for a resonant level with gate dissipation \cite{ChungPRL09}, for instance, but not for the more complex two channel Kondo-like model that we are dealing with here. 

With the approximation $R(\epsilon,T,V)  \approx R(\epsilon,T,V\!=\!0)$, the non-linear current therefore reads, 
\begin{equation}
I(V,T) \approx \frac{e}{h} \int^{D_0}_{-D_0} d\epsilon \ \left[1 - R(\epsilon,T)\right]\left[f_R(\epsilon) - f_L(\epsilon)\right],
\label{rg_conductance}
\end{equation}
where $f_{L/R} (\epsilon)$ is the  Fermi-Dirac distribution. The normalized reflection probability $R(V,T)/R(0,T)$ from \eqref{rg_conductance} exhibits a crossover from  power-law behavior in $V/T$, 
\begin{equation}
R(V,T)/R(0,T) \approx (V/T)^{2/(1+r)}
\quad \text{for }\; V/T > 1, 
\end{equation}
to the constant value $1$ for $V/T \!\to\! 0$, as expected from general considerations (see end of Sec.\,\ref{sec:qc_regime}). 

The result of solving \eqref{rg_conductance} is plotted in 
Fig.\,\ref{fig:S1} (black line) 
and compared to the full theory [Eq.\,\eqref{eq:full_I-V}, red line] as well as experimental data. (See \cite{SupMat} for comparison to $r\!=\!0.75$ data.)
While the correct power-law behavior is captured by this approximation, the magnitude of the conductance (i.e.\ the prefactor) and the cross-over from weak to strong bias are not. The explicit bias dependence of the reflection probability $R(\epsilon,T,V)$ 
\cite{RoschNoneqKondoPRL03,RoschNoneqKondoJPSJ05} clearly would be essential in moving the RG curve toward the experimental data and the full theory.

\vspace*{-0.1cm}
\section{Interpretation as Dynamical Coulomb Blockade}
\label{sec:derriveDCB}

To enhance the physical understanding of our main results, Eq.\,\eqref{eq:full_I-V} and Fig.\,\ref{fig:GV}, we rewrite our strong-hybridization effective system as a fermionic problem and thereby make a direct connection to dynamical Coulomb blockade (DCB) theory. In order to use non-interacting fermions, we choose to  refermionize the non-interacting bosonic fields $(\theta_f,\phi_c)$ in Eq.\,\eqref{eq:Hboson}, using the bosonization relation Eq.\,\eqref{eq:bosonization} where $\alpha$ now denotes this pair. It is also convenient to move the bias out of the barrier term by undoing the time-dependent gauge transformation. The back-scattering term Eq.\,\eqref{eq:Hboson-backscatter} is, then, replaced by the two terms 
\begin{equation}
\begin{aligned}
H_\textrm{T} & = 
\pi a_0 A \cos\left[2 \sqrt{\pi} \theta_c^{\prime}(0) \right] 
\left[ \psi^\dagger_L(0) \psi_R(0)+ \textrm{h.c.} \right] \\
H_\mu &= -eV \theta'_c(0)/\sqrt{4\pi} .
\end{aligned}
\end{equation}
The second equation here follows from the first term in \eqref{bias_effect} by carrying out the integral. 


The fact that the bias couples to the interacting field $\theta_c^{\prime}$ is a serious complication. However, note that we will calculate the $I$-$V$ curve only to leading order in $A$, as in the main text. In the expression for the rate, the bias appears only in the energy-conservation $\delta$-function as the particle gains (or loses) energy $eV$ when it backscatters. Note that the excitations of $\theta_c^{\prime}$ and $\theta_f$ are tightly linked in the single term in 
Eq.\,\eqref{eq:Hboson-backscatter}, leading to a single connection between a given $|i \rangle$ and its $|f\rangle$. Thus, whether the energy $eV$ comes from coupling to the interacting or non-interacting field cannot be distinguished at this order. We can, then, calculate the $I$-$V$ curve using the bias term  
\begin{equation}
H'_{\mu} = -eV\theta_f(0)/\sqrt{4\pi}.
\end{equation}
Refermionizing this term using relations analogous to those in Eqs.\,\eqref{eq:bosonization} and \eqref{bias_effect}, we arrive at the auxiliary model 
\begin{equation}
\begin{aligned}
\!\!\!H'  &= \frac{1}{2} \int_{-\infty}^{\infty} dx \left[ \psi^{\dagger}_R(x) \partial_x \psi_R(x) - \psi^{\dagger}_L(x) \partial_x \psi_L(x) \right]\\
 & + \frac{1}{2} \int_{0}^\infty dx 
 \Big[ (1+r)(\partial_x {\theta}_c^\prime)^2
+ \frac{1}{1+r} (\partial_x {\phi}_f^\prime)^2 \Big] \\
 & + \pi a_0 A \cos\left[2 \sqrt{\pi} \theta_c^{\prime}(0)\right] \big\{ \psi^\dagger_L(0) \psi_R(0)+ \textrm{h.c.} \big\} \\
 & + \frac{eV}{2} \! \left[  \int_{-\infty}^0 \!\!\!\!\! dx \, \psi^{\dagger}_R(x) \psi_R(x) - \int_0^{\infty} \!\!\!\!\! dx\, \psi^{\dagger}_L(x) \psi_L(x) \right].
\end{aligned}
\label{hst_after_refermionization}
\end{equation}
Each line of Eq.\,\eqref{hst_after_refermionization} can be interpreted physically: the first line is right- and left- moving non-interacting fermions, second line is an interacting bosonic environment, third line shows that backscattering of the fermions excites the environment, and fourth line accounts for the voltage bias between the right- and left- moving fermions. 

We thus recognize the form for tunneling of non-interacting particles through a barrier in the presence of an environment \cite{IngoldNazarov92, NazarovBlanterBook, DevoretEsteveUrbina95}, albeit with a strange barrier and strange environment. Tunneling through the barrier consists of backscattering between two chiral fermion modes, and the environment $\theta_c'$ involves a nonlinear combination of the original electrons and environment.
Nevertheless, the standard techniques of DCB theory \cite{IngoldNazarov92, NazarovBlanterBook, DevoretEsteveUrbina95} can be applied to obtain the nonlinear $I$-$V$ curve to second order in $A$.

More specifically, with the Hamiltonian Eq.\,\eqref{hst_after_refermionization} we calculate the tunneling rate through the dynamical Coulomb blockade method
\begin{equation}
\begin{aligned}
&\Gamma(V,T)\\ &=\frac{2 \pi}{\hbar} \int^{+\infty}_{-\infty} 
\!\!\!\! dE^i dE^f \sum_{R^i_1 R^f_1} |\langle E^i | H_r^F | E^f \rangle  |^2 | \langle  R^i | H_r^B | R^f \rangle |^2  \\
& \times P_{\beta} (R^i) P_{\beta}(E) \delta (E^i+E_R^i+e V-E^{f}-E_R^f),
\end{aligned}
\label{eq:DCBgamma}
\end{equation}
where $| E^{i} \rangle$ represents the initial state of a quasi-particle ($\psi^{\dagger}_R$) in the right-moving channel with energy $E^i$ and $| E^f \rangle$ refers to the left-moving final state.
Their statistics is described by the fermionic distribution $P_{\beta}(E)$.
On the other hand, $\theta'_c$ now functions as the dissipative environment, with its initial and final states $| R^{i,f} \rangle$ and the initial bosonic distribution function $P_{\beta}(R^i) = \langle R^i | \rho_{\beta} | R^i \rangle$ (here $\beta$ is a reminder that the density of states is thermally dependent).

With the help of the delta function identity Eq.\,\eqref{delta_fourier}, 
Eq.\,\eqref{eq:DCBgamma} becomes time dependent and can be separated into two parts,
one that involves the bosonic environment and the other the fermionic particles. 
For the bosonic part, we find 
\begin{widetext}
\begin{equation}
\begin{aligned}
 &\sum_{R^i , R^f} \!\! |\langle R^{f} | \cos\left[ 2 \sqrt{\pi} \theta^{\prime}_c(0) \right] | R^i \rangle |^2 \cdot e^{\frac{i}{\hbar} (E_{R}^i-E_{R}^{f}) t} P_{\beta}(R^i) \\
=&  \sum_{R^i , R^f} \!\! \langle R^i | \cos \left[ 2 \sqrt{\pi} \theta^{\prime}_c(t, 0) \right] | R^{f} \rangle 
\langle R^{f} | \cos \left[ 2 \sqrt{\pi} \theta^{\prime}_c(0, 0) \right] | R^i \rangle P_{\beta}(R^i) \\
 =& \left\langle \cos\left[ 2 \sqrt{\pi} \theta'_c(t,0)\right] \cos\left[ 2 \sqrt{\pi} \theta'_c(0,0)\right] \right\rangle =\frac{1}{4} \left\langle e^{i 2 \sqrt{\pi} \theta^{\prime}_c (t,0)} e^{-i 2 \sqrt{\pi} \theta^{\prime}_c (0,0)} \right\rangle  =  \frac{1}{4} e^{J(t)} ,
\end{aligned}
\label{sandwiched}
\end{equation}
\end{widetext}
where $J(t) \equiv 4\pi\langle \left[ \theta'_c (t) - \theta'_c (0) \right] \theta'_c (0) \rangle$ is the phase-phase correlation function. 
[In obtaining Eq.\,\eqref{sandwiched}, we used the relations \cite{IngoldNazarov92}  $\langle e^{i \theta(t)} e^{i \theta(0)} \rangle = 0$ and
$\langle e^{i 2 \sqrt{\pi} \alpha \theta(t)} e^{-i 2 \sqrt{\pi} \alpha \theta(0)} \rangle = e^{ \alpha^2 4 \pi \langle \left[ \theta(t) - \theta(0) \right] \theta(0) \rangle}=e^{\alpha^2 J (t)}$.]
Since the free bosonic action is quadratic, we can calculate this correlation with a Gaussian integral \cite{IngoldNazarov92,SassettiWeissEPL94}
\begin{equation}
\begin{aligned}
J (t) & = -\frac{2}{1+r} \ln \sinh \left(\frac{\pi k_B T |t|}{\hbar}\right) +\frac{2}{1+r} \ln \frac{\pi k_B T }{\hbar \omega_R} \\
& \ \ \ \ - \frac{2}{1+r} \frac{i \pi}{2} \text{Sign} (t) - \frac{2}{1+r} \gamma,
\label{jt_original}
\end{aligned}
\end{equation}
where $\omega_R$ is the energy cutoff of the bosonic bath and $\gamma$ is Euler's constant.

Next we deal with the fermionic part. In the DCB method \cite{IngoldNazarov92}, the backscattering barrier is treated as an effective backscattering resistance $R_T$ so that the fermionic matrix element can rewritten as
\begin{equation}
\left| \langle E^i | H_r^F | E^f \rangle \right|^2 P_{\beta}(E) = \frac{\hbar}{2 \pi e^2 R_T} f(E^i) \big[1 - f(E^f)\big],
\label{fermionic_matrix_element}
\end{equation}
where $f(E)$ represents the equilibrium Fermi-Dirac distribution.

Combining the fermionic and bosonic parts and including the phase factor $\exp\left[i (E^i -E^f +eV) t/\hbar\right]$, we arrive at the expression for the backscattering rate
\begin{equation}
\begin{aligned}
\Gamma(V,T) &\! = \! \frac{1}{2 \pi \hbar e^2 R_T}\!\! \int_{\infty}^{\infty} \!\!\!dE^i dE^f f(E^i) \left[1 \!-\! f(E^f + eV)\right] 
\\
& \qquad\qquad\qquad \times \int_{-\infty}^{+\infty} \!\!dt e^{J(t)} e^{\frac{i}{\hbar} (E^i - E^f) t} \\[0.3cm]
 & = \frac{\hbar \omega_R}{2 \pi e^2 R_T} \frac{e^{\frac{e V}{2 k_B T}}}{\mathbf{\Gamma}(\frac{2}{1+r} + 2)} 
\left(\frac{2 \pi k_B T}{\hbar \omega_R}\right)^{\frac{2}{1+r}+1} \\ 
&\ \ \ \  \times 
\left|\mathbf{\Gamma}\left(\frac{1}{1+r} +1 + i\frac{e V}{2 \pi k_B T}\right)\right|^2,
\end{aligned}
\label{sacttering_rate}
\end{equation}
where $\mathbf{\Gamma}(x)$ is the Gamma function. 
Physically, this rate only involves tunneling from the right-moving to left-moving channel. The net tunneling rate is described by the difference $\Gamma(V,T) - \Gamma(-V, T)$. Since the energy associated with the bias in each backscattering process is $eV$, we can reasonably argue that the charge carried by each quasi-particle is $e$. Consequently, the backscattering-related current is $\Delta I(V,T) = e\left[\Gamma(V,T) - \Gamma(-V,T)\right]$.

As a limiting case, we know from Eq.\,\eqref{hst_after_refermionization} that when $A = 0$ the two fermionic chiral channels are decoupled, and the system conducts perfectly, $G=e^2/h$. Thus we conclude that the current is 
\begin{equation}
\begin{aligned}
I(V,T) & = \frac{e^2}{h} V - \Delta I(V,T) \\
& = \frac{e^2}{h} V - e \left[\Gamma(V,T) - \Gamma(-V, T)\right] \\
 & =\frac{e^2}{h} V - \frac{V}{R_T} \frac{1}{\mathbf{\Gamma}(\frac{2}{1+r}+2)} \left(\frac{2 \pi k_B T}{\hbar \omega_R}\right)^{\frac{2}{1+r}} 
\\
& \ \ \ \ \ \ \times \frac{|\mathbf{\Gamma}(\frac{1}{1+r}+1+i\frac{e V}{2 \pi k_B T})|^2}{|\mathbf{\Gamma}(1+i\frac{e V}{2 \pi k_B T})|^2},
\end{aligned}
\label{current_DCB}
\end{equation}
where we have used the identity $\sinh(\pi x) \!= \pi x 
|\mathbf{\Gamma}(1+i x)|^2$.

The result Eq.\,\eqref{current_DCB} 
is the same as the current expression in the main text, Eq.\,\eqref{eq:full_I-V}, with the coefficient of the correction $(A\pi/\hbar\omega_R)^2$ replaced by $(h/e^2)/R_T$, where $R_T$ is the tunneling resistance of the effective barrier in the absence of dissipation. 

The absolute magnitude of the current in the power-law scaling regime is a key prediction of weak-coupling (large barrier) DCB theory. Here, because of the unknown amplitude $A$ in the effective strong-hybridization model Eq.\,\eqref{eq:Hboson} or \eqref{hst_after_refermionization}, we do not have such a prediction. However, for the normalized quantity plotted in the figures, DCB theory does give a definite value of the prefactor because it is fixed by the way the theory transitions from the crossover to the asymptotic regime. It is exactly this prefactor and crossover that is missed in the approximate RG theory outlined in Appendix~\ref{sec:RGforIV},
thus highlighting the importance of additional dephasing processes not included there but fully included in the calculation in 
Sec.\,\ref{sec:I-Vcurve} 
as well as in the DCB calculation here.  


The equivalence of these two coefficients is shown by considering the standard single-barrier tunneling Hamiltonian. Denote the backscattering amplitude of the fermions by $t_{k,q}$, where $k$ and $q$ label the initial and final fermionic particle states, $H_T = \sum_{k,q} t_{k,q} c^\dagger_{L,k} c^{\,}_{R,q} + \text{h.c.}$. 
The standard result for the conductance of a tunneling barrier when the amplitude is momentum independent is 
$1/R_T = (e^2/h) |\overline{t}|^2 [\Xi N(0)]^2$, where $\Xi N(0)$ is the number of states per unit energy and $\overline{t}$ is the average matrix element. In our case, the number of states is the size of the system $L$ divided by the bosonization cutoff $a_0$, and the maximum energy for a particle excitation is $\hbar \omega_R$, the cutoff for the bosonic modes ($-\hbar \omega_R$ for a hole excitation). The amplitude $\overline{t}$ follows from Eq.\,\eqref{hst_after_refermionization} noting that a factor of $1/L$ is introduced in the conversion from continuous $x$ to discrete $k$. Putting these elements together one finds
\begin{equation}
\frac{1}{R_T} = \frac{e^2}{h} \left(\frac{\pi a_0 A}{L}\right)^2 \left(\frac{L/a_0}{\hbar \omega_R}\right)^2
= \frac{e^2}{h} \left(\frac{\pi A}{\hbar \omega_R}\right)^2 .
\end{equation}
Thus the $I$-$V$ curve that results from a DCB theory treatment of the auxiliary strong-hybridization model Eq.\,\eqref{hst_after_refermionization} and that found from the true effective bosonic description Eq.\,\eqref{eq:Hboson} are \emph{identical}.


\emph{This allows then the physically intuitive interpretation of the $I$-$V$ curve Eq.\,\eqref{eq:full_I-V} as tunneling of non-interacting fer\-mions (between left-movers and right-movers) in the presence of an environment.}

\newpage
\bibliography{QTransport_2021-01,BibFootnotes}

\begin{thebibliography}{114}
\bibitem [{Bia()}]{BiasLL}%
  \BibitemOpen
  \href@noop {} {}\bibinfo {note} {The bias-dependent factor $eVt$ is not
  modified by the interaction after including the effect of the boundary
  between the 1D wire and the source and drain reservoirs, as is standard in 1D
  quantum transport
  \cite{PonomarenkoRPB95,SafiSchulzPRB95,MaslovStonePRB95,FrohlichPRB96,HuKaneX16}.
  See the discussion in the last paragraph of
  Appendix\,\ref{sec:integrateout}.}\BibitemShut {Stop}%
\bibitem [{Per()}]{PerfectTransmission}%
  \BibitemOpen
  \href@noop {} {}\bibinfo {note} {To explicitly obtain a translationally
  invariant system at the QCP, first rescale $\theta_c'$ and $\phi_f'$ in
  Eq.\,\eqref{eq:Hboson-freeint} so as to absorb the factors of $(1+r)^{\pm1}$.
  When $A\!=\!0$, this can be done without any compensating change in other
  terms. Then combine the flavor and charge fields into fields defined for all
  $x$, e.g.\ $\theta'(x\!>\!0)=\theta_c'(x)\!+\!\theta_f(x)$,
  $\theta'(x\!<\!0)=\theta_c'(-x)\!-\!\theta_f(-x)$, and similarly for
  $\phi(x)$. The Hamiltonian now becomes $\int_{-\infty}^\infty dx
  \big[(\partial_x {\theta'})^2 + (\partial_x {\phi'})^2 \big]$, which is
  manifestly translationally invariant.}\BibitemShut {Stop}%
\bibitem [{Fra()}]{FractionalCharge}%
  \BibitemOpen
  \href@noop {} {}\bibinfo {note} {Strictly speaking, the quasi-particles carry
  fractional charge in the presence of dissipation. Experimentally, however,
  the fractional charge becomes undetectable when the dissipative system
  couples to two metallic reservoirs. This effect is similar to that in
  Luttinger liquid systems \cite{PonomarenkoRPB95,SafiSchulzPRB95,
  MaslovStonePRB95, FrohlichPRB96, FendleyPRB95, HuKaneX16}.}\BibitemShut
  {Stop}%
\bibitem [{Imp()}]{ImpurityEntropy}%
  \BibitemOpen
  \href@noop {} {}\bibinfo {note} {At the QCP, certain degrees of freedom
  associated with the dot become decoupled and contribute a boundary entropy.
  The best known example is the $\ln(\sqrt{2})$ entropy at the Emery-Kivelson
  point of the two-channel Kondo model due to the decoupled Majorana fermion.
  In our case, the boundary entropy can be calculated from conformal field
  theory to be $\ln(\sqrt{1+r})$
  \cite{WongAffleck94,Zheng1-GPRB14}.}\BibitemShut {Stop}%
\bibitem [{QBM()}]{QBM}%
  \BibitemOpen
  \href@noop {} {}\bibinfo {note} {The effective action of our problem maps to
  that of 1D quantum Brownian motion \cite{YiKanePRB98}. Ref.\ \cite{WeissBook}
  discusses the duality between weak and strong coupling in that case (Section
  26, ``Duality Symmetry'').}\BibitemShut {Stop}%
\end{thebibliography}


\newpage
\global\long\def\thefigure{S\arabic{figure}}
\setcounter{figure}{0}

\begin{center}
\textbf{Supplemental Material for \\
      \vspace{5pt}
``Nonequilibrium Quantum Critical Steady State: Transport Through a Dissipative Resonant Level''}\\
      \vspace{10pt}
Gu Zhang, Chung-Hou Chung, Chung-Ting Ke, \\C.-Y. Lin, Henok~Mebrahtu, Alex I. Smirnov,
\\Gleb Finkelstein, and Harold U. Baranger\\
\vspace{5pt}
{(26 January 2021)}
\end{center}
\vspace*{1.1cm}

\begin{figure}[h]
\includegraphics[width=0.93\columnwidth]{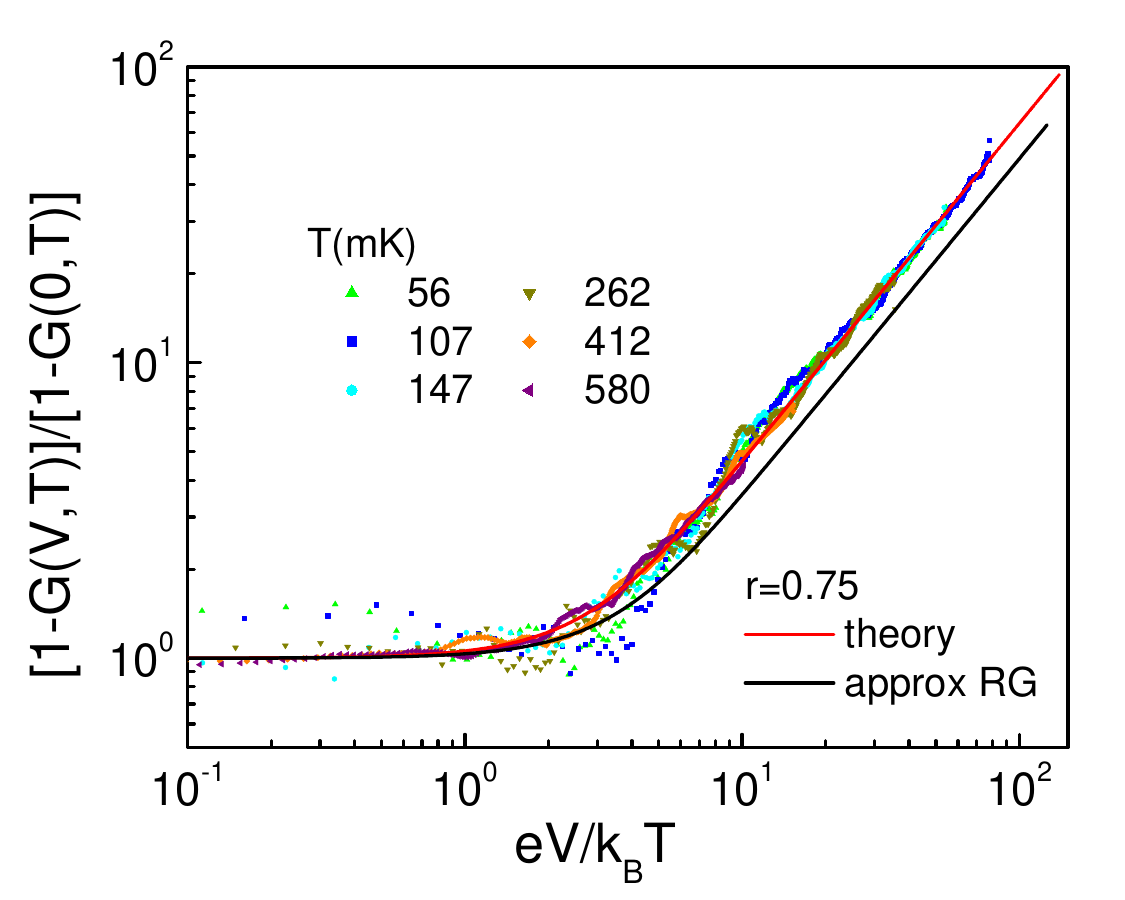}
\vskip -0.3cm
\caption{
Deviation from perfect conductance $1-G(V,T)$, scaled by $1-G(V\!=\!0,T)$, as a function of $eV/k_B T$ for $r \!=\! 0.75$. [$G(V,T)$ is the differential conductance, $G\!\equiv\!(h/e^2)\,dI/dV$.]  
The symbols are experimental results at the color-coded temperatures. The red and black lines result from the full non-equilibrium and approximate RG theories, respectively, in which there are no free parameters [Eqs.\,\eqref{eq:full_I-V} and \eqref{rg_conductance}, respectively]. As for the $r=0.5$ data shown in the main text, note the excellent agreement between the full theory and the data in both the crossover and power-law regimes. The deviations from scaling seen at high bias in Fig.\,\ref{fig:GV} of the main text were not investigated in this sample as the bias range was limited.  
The experimental data are taken from Ref.\,\cite{Mebrahtu13}.
}
\label{GVp75}
\end{figure}

To confirm the experimental features highlighted in the main text, in Fig.\,\ref{GVp75} we compare data and theory for $r\!=\!0.75$. The agreement between the full theory (red line) and the experiment is excellent. Note in particular that the theory goes right through the data in both the crossover region from low to high bias and the scaling regime. Thus our theory goes well beyond the frequently used scaling arguments that produce only the exponent in the scaling regime (the slope on this log-log plot) and not the actual conductance magnitude. 

As an example of a naive scaling argument, the approximate RG curve is taken from the argument in Appendix~\ref{sec:RGforIV}. As noted there, while the correct power-law behavior is captured by this approximation, the magnitude of the conductance (i.e.\ the prefactor) and the cross-over from weak to strong bias are not. The error in this argument comes from the neglect of the explicit bias dependence of the renormalization.

\begin{figure}[h]
\includegraphics[width=0.85\columnwidth]{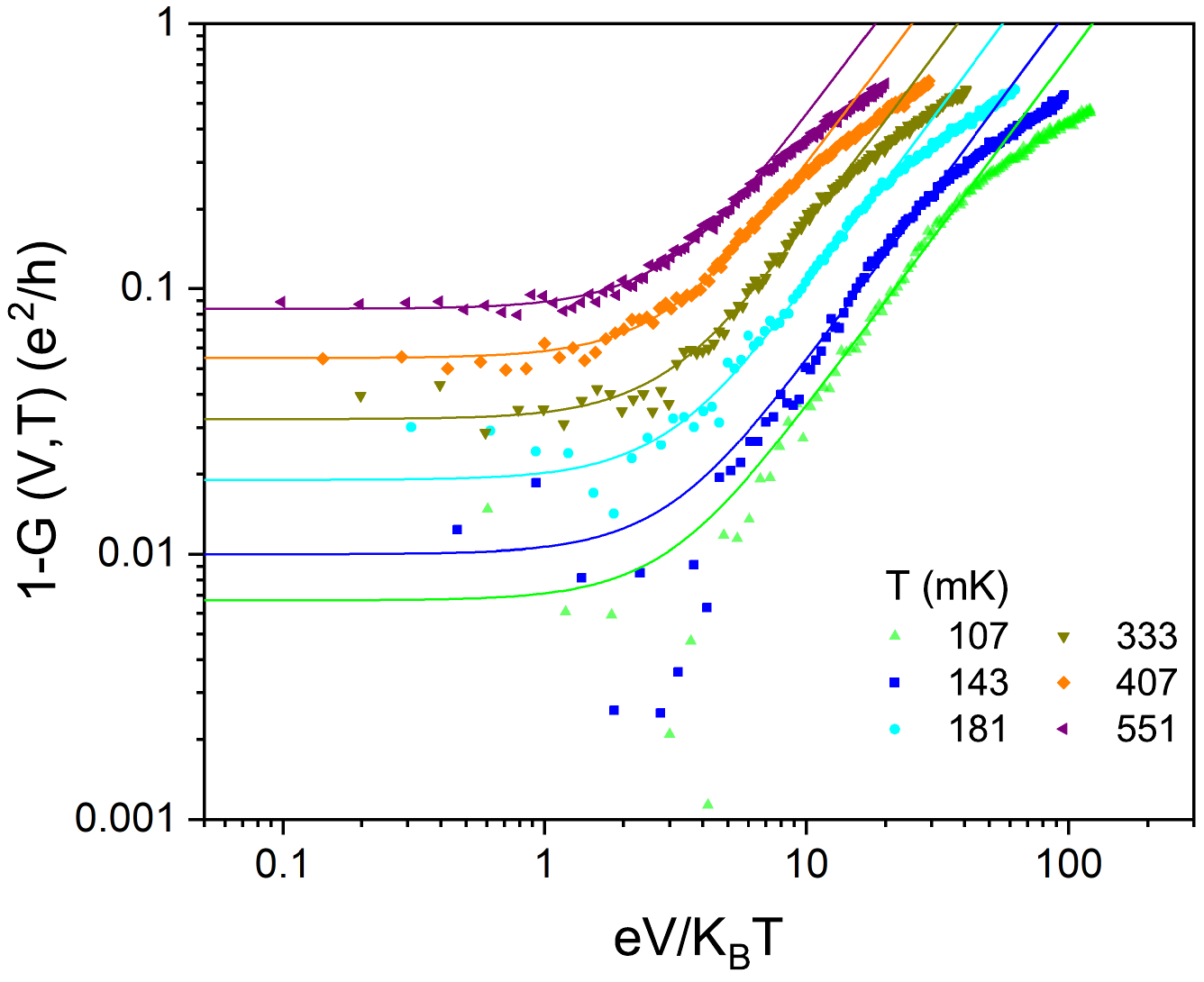}
\vskip -0.1cm
\caption{\label{fig:S3}Unnormalized experimental data points and full theoretical curves for the $r=0.5$ case on a log-log scale (same data as in in Fig.\,\ref{fig:GV} of main text). There is one parameter in fitting each theory curve to the data. 
}
\end{figure}

In addition, it may be valuable to see the experimental data unnormalized. This allows the reader to see, for instance, at what absolute value of the conductance the $I$-$V$ curve becomes non-universal. The corresponding theoretical curves must now, of course, be fit (one parameter). The same data as in Fig.\,\ref{fig:GV} of the main text, namely for $r\!=\!0.5$, is shown in Fig.\,\ref{fig:S3}. 

\newpage 

\end{document}